\documentclass[aps,prb,twocolumn,epsfig,showkeys]{revtex4}
\usepackage[sort&compress]{natbib}
\usepackage
{
  amsmath,
  appendix,
  array,
  epsfig,
  color,
  hyperref,
  latexsym,
  float,
  subfigure,
  longtable,
  wrapfig
}

\newcommand{\eps}{\epsilon}
\newcommand{\dgr}{\dagger}

\newcommand{\D }{\Delta }
\newcommand{\up}{\uparrow}
\newcommand{\dwn}{\downarrow}

\begin{document}

\title{Non-collinear Magnetic Order in the Double Perovskites:\\
  Double Exchange on a Geometrically Frustrated Lattice}

\author{Rajarshi Tiwari and Pinaki Majumdar}

\affiliation{
  Harish-Chandra  Research Institute,\\
  Chhatnag Road, Jhusi, Allahabad 211019, India
}

\date{30 April 2011}

\begin{abstract}
Double perovskites of the form A$_2$BB$'$O$_6$ usually involve a 
transition metal ion, B, with a large magnetic moment, and a non 
magnetic ion B$'$.  While many double perovskites are ferromagnetic, 
studies on the underlying model reveal the possibility of antiferromagnetic 
phases as well {\it driven by electron delocalisation.} In this paper 
we present a comprehensive study of the magnetic ground state and $T_c$ 
scales of the minimal double perovskite model in three dimensions using 
a combination of spin-fermion Monte Carlo and variational calculations. In 
contrast to two dimensions, where the effective magnetic lattice is bipartite, 
three dimensions involves a geometrically frustrated face centered cubic (FCC) 
lattice. This promotes non-collinear spiral states and `flux' like phases 
in addition to collinear anti-ferromagnetic order.  We map out the possible 
magnetic phases for varying electron density, `level separation' $\epsilon_B - 
\epsilon_{B'}$, and the crucial B$^{\prime}$B$^{\prime}$
(next neighbour) hopping $t'$. 
\end{abstract}


\maketitle

\section{Introduction}

Double perovskites (DP) constitute a large family of materials 
\cite{dp-rev} with
molecular formula A$_2$BB$'$O$_6$, where A is an alkali or alkaline
earth metal, and B and B$'$ are typically transition metals.
Although double perovskites have been studied for 
decades~\cite{Anderson1993197}, the discovery of
high $T_c$ ferromagnetism and half-metallicity in Sr$_2$FeMoO$_6$ 
has led to renewed interest in their properties. Later, in a number of
explorations it was discovered that these materials are 
candidates for various technological applications, 
{\it e.g.}, in  
spintronics~\cite{tokura} (Sr$_2$FeMoO$_6$), 
magneto-dielectrics~\cite{rogado:2005,umesh08} (La$_2$NiMnO$_6$),
and magneto-optics~\cite{molly} (Sr$_2$CrOsO$_6$,Sr$_2$CrReO$_6$).
Their properties are determined by the couplings on the
B and B$'$ ions, the B and B$'$ valence state, and the 
structural order in the B-B$'$ lattice. 

The magnetism in the DP's arises from a combination
of $(i)$ Hund's coupling on the B, B$'$ ions and $(ii)$ electron
delocalisation. While there are important DP's where both
B and B$'$  are magnetic ions, in the current work we will
restrict ourselves to materials where only one ion,
B, say, is magnetic.
For example, 
in Sr$_2$FeMoO$_6$ (SFMO) the B atom (Fe) is
magnetic while B$'$ (Mo)  is non magnetic\cite{PhysRevLett.85.2549}.
Even in this restricted class, a large variety of compounds can be
realized by taking $3d$, $4d$ or $5d$ transition metals as B and B$'$, 
and alkaline earths or rare-earths as A. 
These lead to a variety of properties, {\it e.g.}, 
high $T_c$ ferro(or ferri) magnetism (FM),
with half-metallic\cite{topwaldd,majewski05} or insulating 
\cite{krockenberger:020404} behaviour.

\begin{figure}[b]
  \psfig{figure=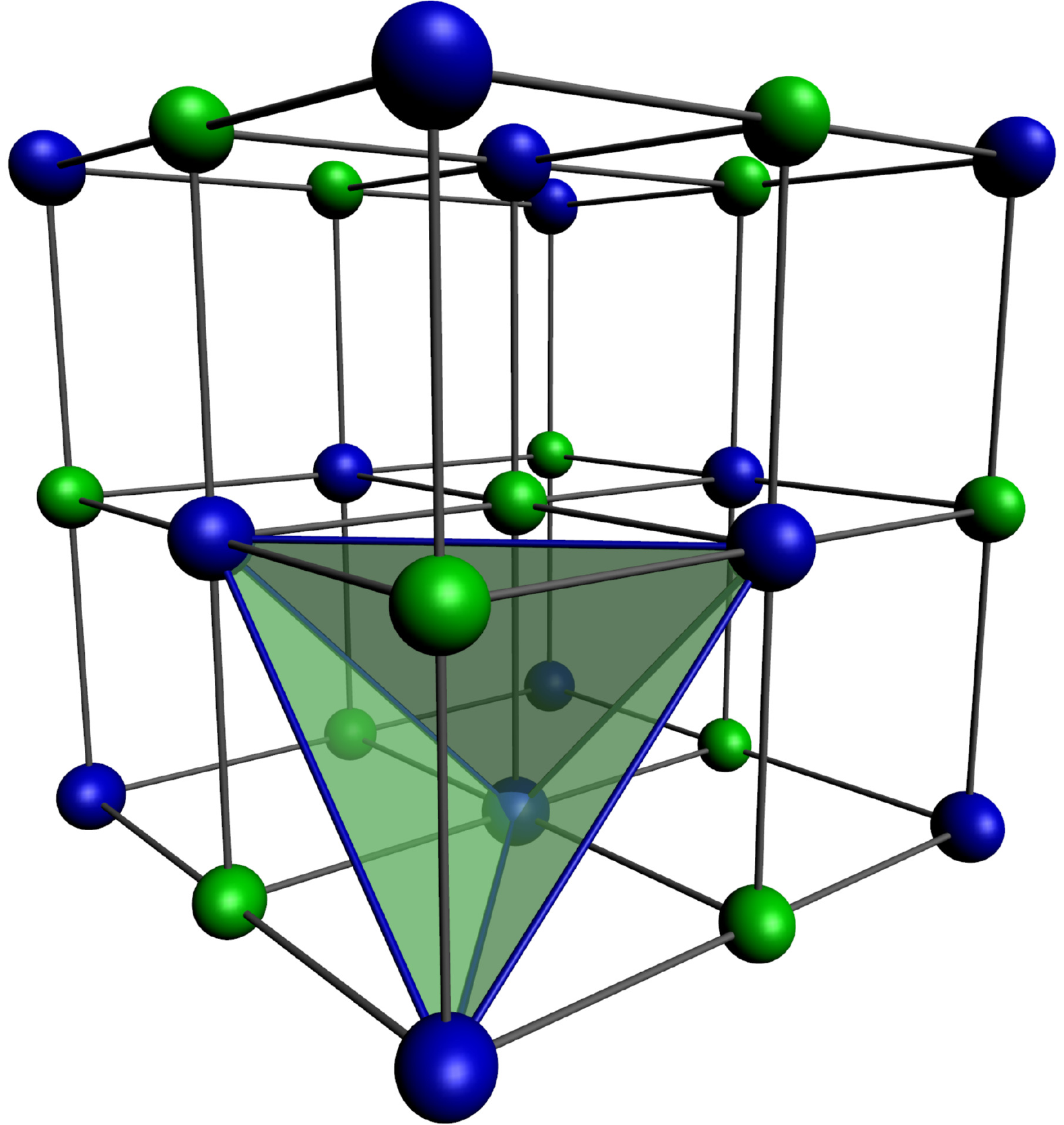,width=6.0cm,height=6.0cm,angle=0}
  \caption
  {\label{b1b2} Colour online: 
    The structure of B-B$'$ lattice in a ordered double
    perovskite. The B and B$^{\prime}$ alternate (as in rock-salt)
    in the ordered structure. If the bottom corner (blue) atom
    is B, then its B nearest neighbours (connected by lines)
    are also nearest neighbours of each other. The triangles
    preclude a `G type' antiferromagnetic phase.}
\end{figure}

There have been several attempts at a theoretical understanding of the
magnetism in these materials. These consist of (i)~{\it ab initio}
electronic structure calculations,  and (ii)~model Hamiltonian based
approaches. The {\it ab initio} 
calculations\cite{umesh08,molly,PhysRevLett.85.2549}
provide material specific information about the electronic structure
and allow a rough estimate\cite{mandal:134431} 
 of the  $T_c$. Unfortunately,
these calculations are rather complicated for non-collinear magnetic
phases that are likely in a frustrated magnetic lattice, see Fig.\ref{b1b2}. 
In such situations model Hamiltonian studies can provide some insight on
possible ordered states.

\begin{figure}[b]
  \psfig{figure=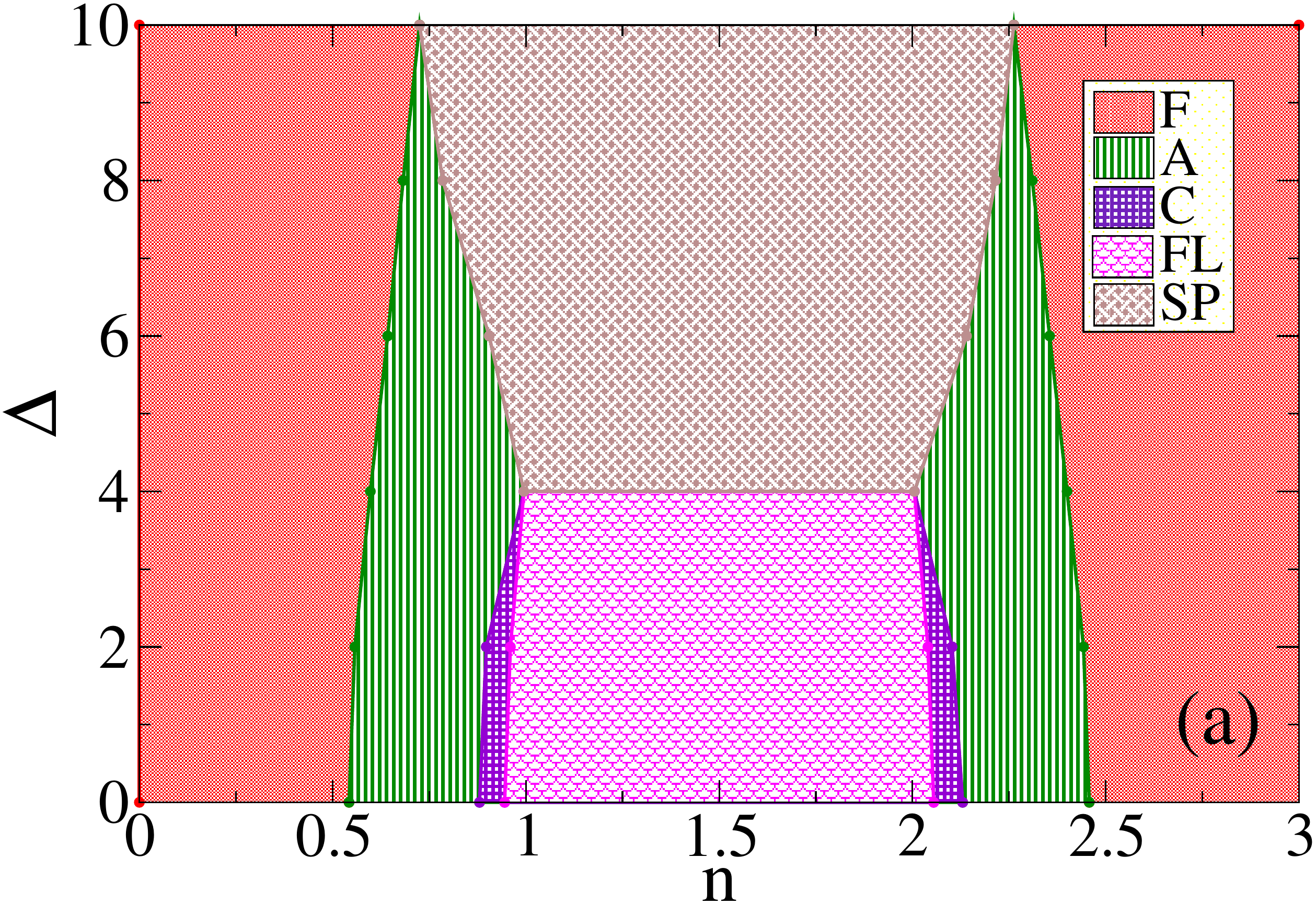,width=6.5cm,height=5.8cm,angle=0}
  \psfig{figure=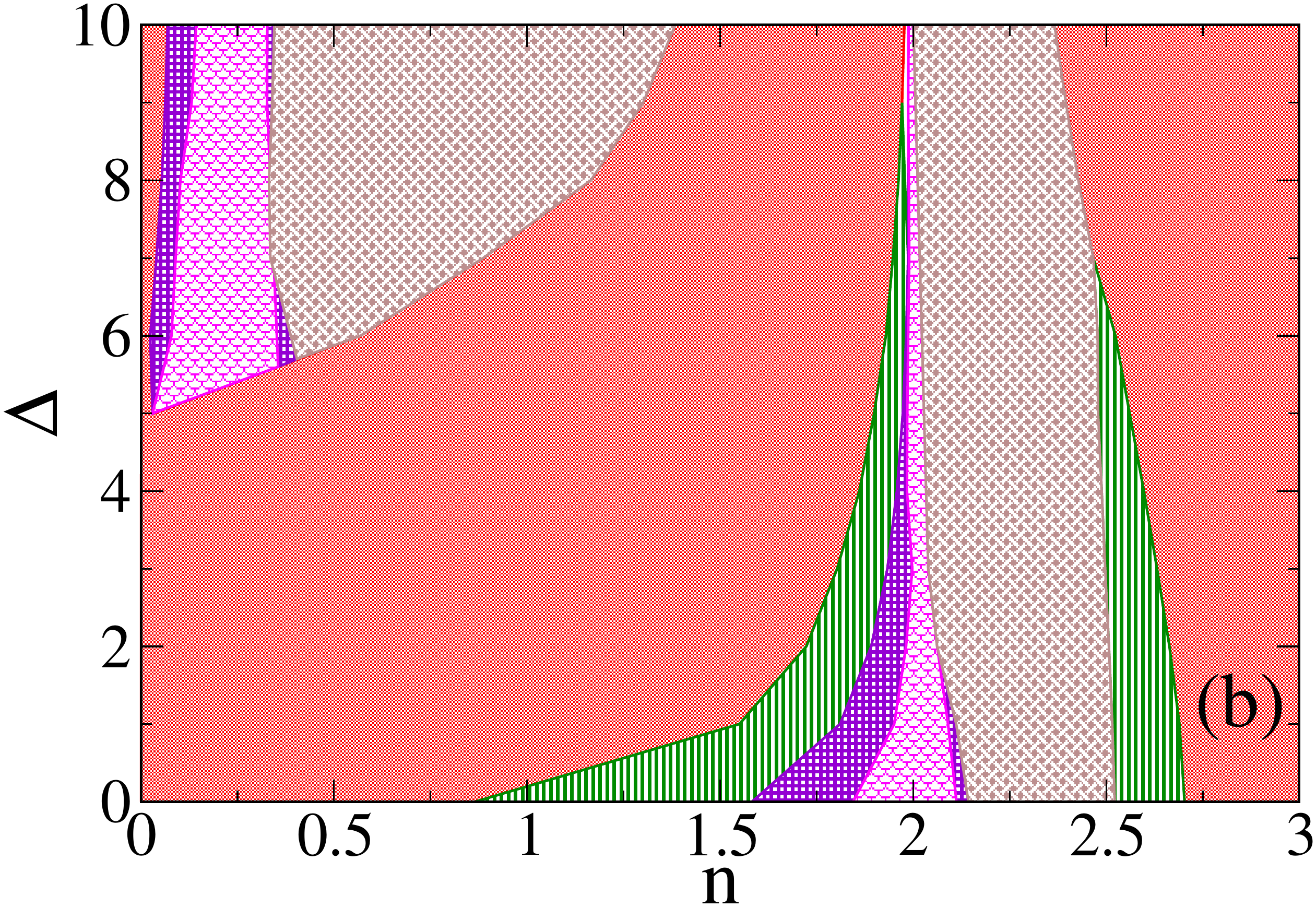,width=6.5cm,height=5.8cm,angle=0}
  \caption
  {\label{phdg1} Colour online: 
Magnetic ground state for varying electron 
density, $n$, and effective B-B$'$ level separation, $\Delta$.
Top: phase diagram with only BB$'$, {\it i.e.},
nearest neighbour, hopping. Bottom: phase diagram when an additional
B$'$B$'$ hopping, $t'/t=-0.3$, is included. The labels are: F (ferromagnet),
A (planar phase), C (line like), FL (`flux') and SP (spiral).
This figure does not show the
narrow windows of phase separation in the model.
The phase diagrams are generated via a combination of Monte Carlo
and variational calculations on lattices of size upto $20 \times
20 \times 20$.
} \end{figure}

\begin{figure}[t]
  \psfig{figure=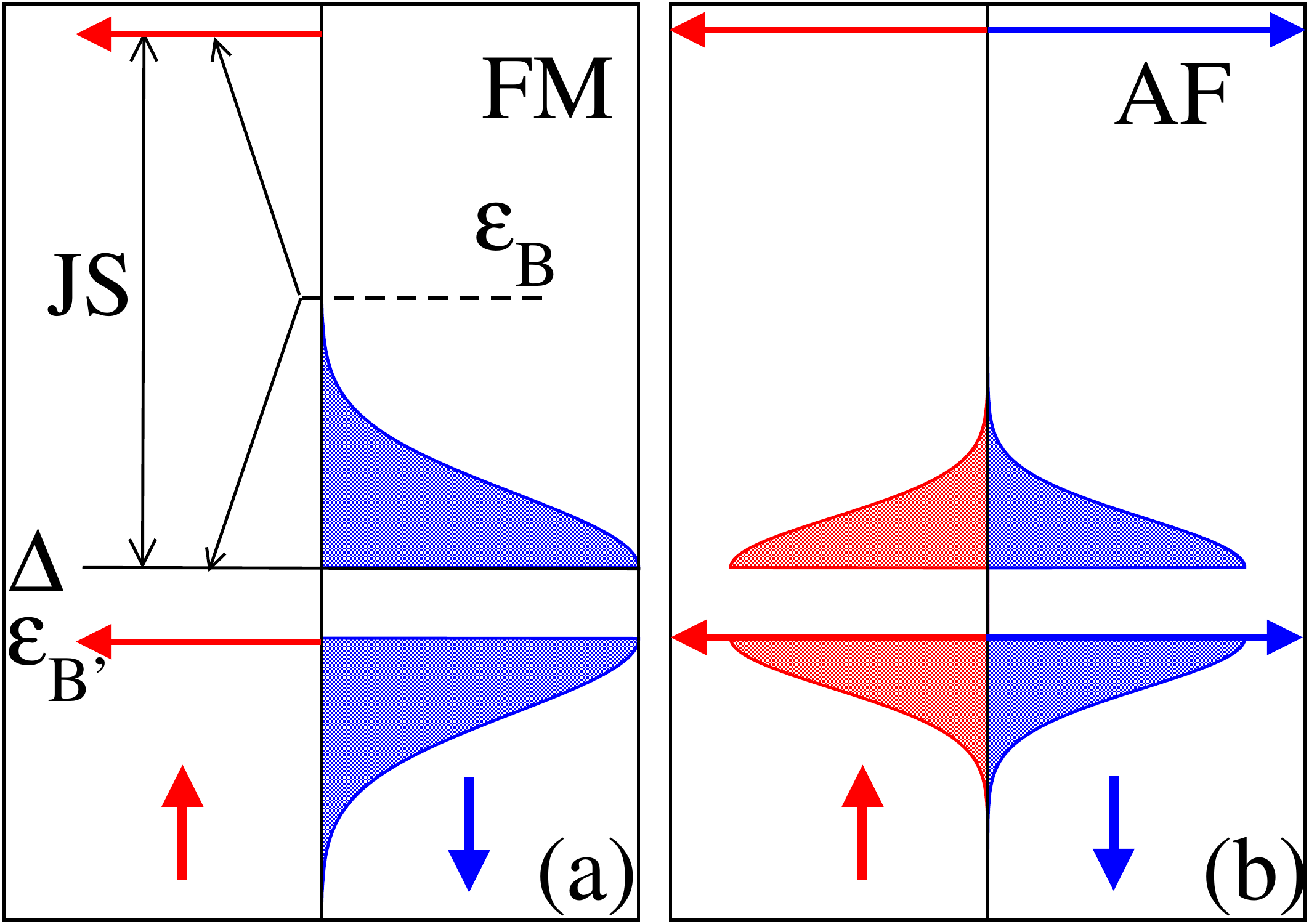,width=7.8cm,height=5.0cm,angle=0}
  \caption
  {\label{vlabel} Colour online: 
Level scheme and schematic band structure 
    for the DP model when only B-B$'$ hopping is allowed.  The 
arrows denote localised atomic levels.
    Red and blue denote $\uparrow$ and $\downarrow$ spins respectively.
The atomic level scheme is shown in (a). where the spin degenerate
B$'$ levels are at $\epsilon_{B'}=0$ and the spin split B levels are
at $\epsilon_B \pm JS/2$. We define the effective B level as
$ \Delta = \epsilon_B -  JS/2$. When $JS \gg t$, the levels at
$\epsilon_{B'}$ and $\Delta$ hybridise to create bands, shown for
the FM case in (a), and for a collinear AF phase in (b). 
  }
\end{figure}

Early model calculations for DP's  used dynamical mean
field theory (DMFT) to estimate $T_c$ and the magnetic stability 
window~\cite{PhysRevB.64.024424,PhysRevB.67.125119}, focusing on ferromagnetism.
Earlier work on the classical Kondo lattice 
\cite{kp-pm-klm,PhysRevB.51.3027,PhysRevB.62.13816,PhysRevB.64.054408}
had revealed that variation in carrier density can lead to a wide variety
of phases in a spin-fermion problem. Indeed, calculation~\cite{sanyal:054411}
in a two dimensional (2D) model of DP's  
confirmed the existence of antiferromagnetic (AF),
albeit collinear, phases. 
The `frustrated' character of the three dimensional (3D)
DP lattice raises the
intriguing possibility of doping driven 
non collinear magnetic phases as well.
Our study aims to explore this issue in detail.

Our main results are the following. Using a combination of
Monte Carlo and variational minimization, we map out the
magnetic ground state (Fig.\ref{phdg1}) at large Hund's
coupling for varying electron density and B-B$'$ level separation.
In addition to FM, and collinear A and C type order, the phase 
diagram includes large regions of non-collinear `flux'  and spiral
phases and windows of phase separation. Modest B$'$B$'$ hopping leads
to significant shift in the phase boundaries, and ``particle-hole asymmetry''. 
We provide  estimates of the $T_c$ of these non trivial 
magnetic phases.

The paper is organized as follows. In Section~\ref{modmeth}
we describe the model and methods, Section~\ref{ph-sym-case}
describes our results in the particle-hole symmetric
case ($t'=0$), and Section~\ref{ph-asym-case} describes the effect 
of finite $t'$. Section~\ref{discuss}
discusses some issues of modeling the real DP.
Section~\ref{conclude} concludes the paper.

\section{Model and method}
\label{modmeth}
Previous study of double perovskites in two dimensions~\cite{sanyal:224412}
revealed three collinear phases, namely FM, a diagonal stripe 
phase (FM lines coupled antiferromagnetically) 
and a `G type' phase (up spin 
surrounded by down and {\it vice versa}).
In 2D the B sub-lattice is square and 
bipartite, so there is no frustration. In a 3D 
{\it simple cubic} B lattice the counterparts of the 2D
phases would be FM, A type (planar),  C type (line like)
and G type. The magnetic B ion lattice in the DP's is, however, 
FCC which is {\it non-bipartite}, so while one can construct FM and planar 
A type phases, the C type 
phase is modified and the G type phase cannot exist.

Fig.\ref{b1b2} briefly indicates why it is impossible to
have an `up'($\up$) B ion to be surrounded by only
`down'($\dwn$) B ions, {\it i.e.}, the G type arrangement.
Two B neighbours of a B ion are also neighbours of each
other, frustrating G type order.
The suppression of the G type phase,
which occupies a wide window in 2D, requires us to move
beyond collinear phases in constructing the 3D phase diagram.
We will discuss the variational family in Section~\ref{varsch}.

\subsection{Model}
\label{mod}
The alternating arrangement of  B and B$'$ ions in the ordered cubic double
perovskites is shown in Fig.\ref{b1b2}. We use the following one band model
on that structure:
\begin{eqnarray}
H &= & 
\eps_{B}\sum_{i \in B} f^{\dgr}_{i \sigma}f_{i \sigma} 
+ \eps_{B'}\sum_{i \in B'} m^{\dgr}_{i \sigma}m_{i \sigma}
-\mu {\hat N}
\cr
&&~  
  -t\sum_{<ij >} 
f^{\dgr}_{i \sigma} m_{j \sigma} 
+ J\sum_{i \in B} {\bf S}_{i }\cdot 
f^{\dgr}_{i \alpha} {\roarrow \sigma}_{\alpha \beta}
f_{i \beta} 
\nonumber
\end{eqnarray}

The $f^{\dgr}$ correspond to the B ions and the $m^{\dgr}$ to the
B$'$.
$\eps_{B}$ and $\eps_{B'}$ are `onsite' energy on the
B and B$'$ sites respectively, {\it e.g.}, the $t_{2g}$ level
energy of Fe and Mo in SFMO.
$\mu$ is the chemical potential and
${\hat N} = \sum_{i \sigma} (f^{\dgr}_{i \sigma}f_{i \sigma}
+  m^{\dgr}_{i \sigma}m_{i \sigma})$
is the total
electron number operator.
$t$ is the hopping amplitude between
nearest neighbour B and B$'$ ions. We augment this model later to
study first neighbour B$'$B$'$
hopping $t'$ as well. 
$J$ is the (Hund's)  coupling 
between the B core spin and the $f$ conduction electron. 
We will use 
$\vert {\bf S}_i \vert =1$, and 
absorb the magnitude of ${\bf S}$  in $J$. 
${\sigma}^{\mu}_{\alpha\beta}$ are the Pauli matrices.

\begin{figure}[b]
  \psfig{figure=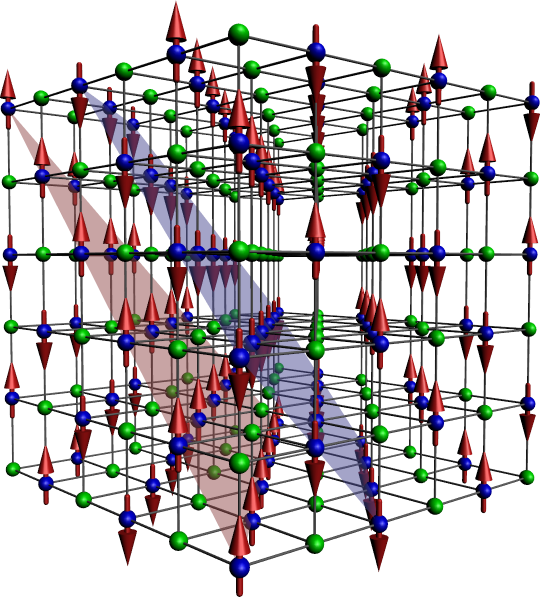,width=6.0cm,height=6.0cm,angle=0}
  \vspace{.2cm}
  \psfig{figure=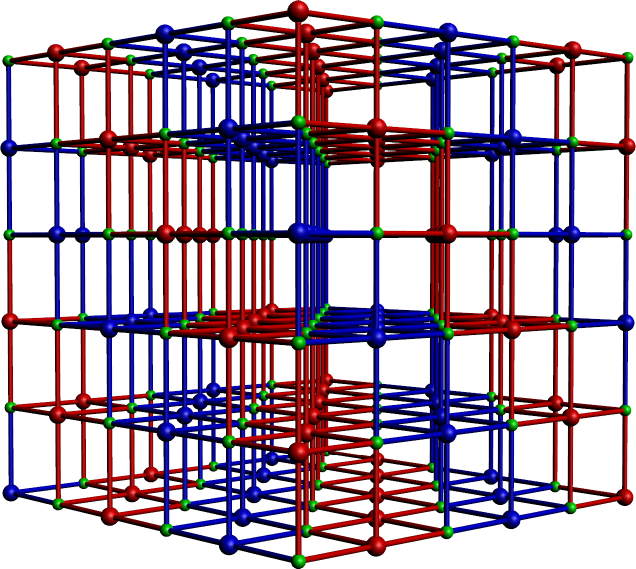,width=6.0cm,height=6.0cm,angle=0}
  \caption{\label{fig:spinA}
    Colour online. Top: Spin configuration for 
    `A type' order. The spins are parallel within the 111 planes 
    (shown) and
    are antiparallel between neighbouring planes.
    Bottom:  The differently coloured bonds show the electron
    delocalisation pathway for up and down spin electrons in the
    A type phase. The delocalisation is effectively two dimensional.}
\end{figure}

\begin{figure}[b]
  \psfig{figure=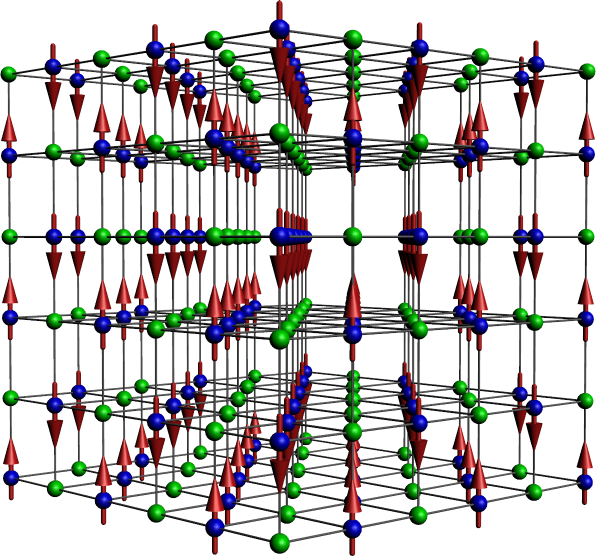,width=6.0cm,height=6.0cm,angle=0}
  \vspace{.2cm}
  \psfig{figure=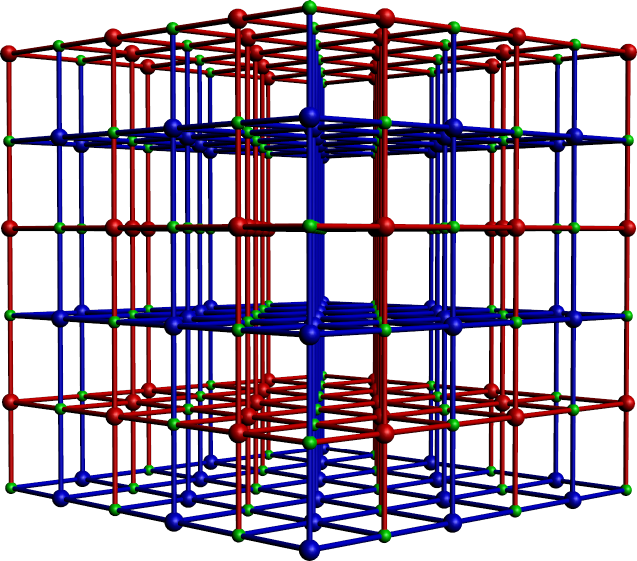,width=6.0cm,height=6.0cm,angle=0}
  \caption{\label{fig:spinC}
    Colour online. Top: Spin configuration in the `C type'
    phase. Core spins are parallel on alternating 110 planes, and 
    antiparallel on neighbouring planes. Bottom: the delocalisation
    path, consisting of the 110 planes and the
    horizontal 001 planes. }
\end{figure}

The model has parameters $J$, $\eps_B$, $\eps_{B'}$, and
$\mu$ (or $n$).
Since only the level difference matters, we
set $\eps_{B'} =0$.
We have set $t=1$, and use $J/t \gg 1$ so that the
conduction electron spin at the B site is slaved to the
core spin orientation. However, to keep the {\it effective}
level difference between B and B$'$ sites finite we use
the parameter $\Delta = \epsilon_B - J/2$, and explore the
phases as a function of $n$ and $\Delta/t$. We will present
results for $t'/t=0$ and $\pm 0.3$.

A schematic for the levels is shown in Fig.\ref{vlabel}.
The structural unit cell of the system has 2 (one B, one  B$'$)
atoms, which amounts to 4 atomic levels (2 up spin, 2
down spin).
The two spin levels at the  B site are separated by $JS$ and
overlap with 2 spin degenerate levels of the
B$'$ site at $\eps_{B'}=0$.
We take the large $J$ limit, and take $\eps_{B} = J/2 + \D$ with $\D$ in
the range (0-10). One B band become centered at $\D$ and second goes
to $JS + \D$. In this situation the down spin B and two B$'$ bands overlap
while up spin B band is always empty. The relevant electron density
window includes the lowest three bands, so our electron density will
be in the range $[0,3]$.

To get a general feel of the band structure of the
particle hole symmetric case, we notice that we have three levels
(excluding the highest $f_\up$ level at $JS +\Delta$ which remains
empty and is redundant for our purpose) in atomic limit.
These include  one spin
slaved $f_\dwn$ level at $\Delta$, and the two $m_\up$, $m_\dwn$,  levels
which overlap with the $f_\dwn$ levels depending on the spin
configurations.
This overlap leads to electron delocalisation and band formation.

For the FM, Fig.\ref{vlabel}.(a), only one spin channel
(say $m_\dwn$) gets to delocalise through $f$ sites and forms two bands,
separated by a band gap of $\Delta$, while other spin channel (say $m_\up$)
is localised at $0$.

For collinear AF configurations,  Fig.\ref{vlabel}.(b),
the conduction path gets divided into two sub-lattices, such that
each spin channel gets to delocalise in one sub-lattice
(in which all the core spins point in same direction, making
the sub-lattice ferromagnetic.) See 
Fig.\ref{fig:spinA}, and Fig.\ref{fig:spinC}
for the details of the conduction path.
In one such sub-lattice, only one
of the $\up$ or $\dwn$ delocalised, the other remains localised. The roles
of $\up$ and $\dwn$ are reversed in going from one sub-lattice to other,
as a result one gets spin-degenerate localised and dispersive bands for
AF phases.

\begin{table*}
  \centering
  \begin{tabular}{|c|cc|}
    \hline
    {\bf Phase} & \multicolumn{2}{c|}{{\bf Peak location in
        $S({\bf q} )$}} \\\hline
    Ferromagnet ({ FM}) & \multicolumn{2}{c|}{(0,0,0),($\pi$,$\pi$,$\pi$)} \\\hline
    A-type antiferromagnet ({ A}) &\multicolumn{2}{c|}{($\frac{\pi}{2}$,$\frac{\pi}{2}$,
      $\frac{\pi}{2}$),($\frac{3\pi}{2}$,$\frac{3\pi}{2}$,$\frac{3\pi}{2}$)}
    \\\hline
    C-type antiferromagnet ({ C}) & \multicolumn{2}{c|}{(0,0,$\pi$),($\pi$,$\pi$,0)} \\\hline
    `flux' ({ FL}) & \multicolumn{2}{c|}{$(\pi,0,0),(0,\pi,0),(0,0,\pi),(\pi,\pi,0),(\pi,0,\pi),(0,\pi,\pi)$}
    \\\hline
    $\up\up\dwn\dwn$ phase & \multicolumn{2}{c|}{$(\frac
      {\pi}{2},0,0),(\frac {3\pi}{2},0,0),(\frac {\pi}{2},\pi,\pi),(\frac{3\pi}{2},\pi,\pi)$} \\\hline
    {\bf SP$_1$} $({\bf q}_{\theta},{\bf q}_{\phi})=(0,\frac{\pi}{2},\pi),(0,0,0)$ &
    $(0,\frac{\pi}{2},\pi),(\pi,\frac{\pi}{2},0),(0,\frac{3\pi}{2},\pi),(\pi,\frac{3\pi}{2},0)$&
    % \qquad\qquad\qquad\qquad\qquad\qquad\qquad\qquad\qquad\qquad\qquad
    \\\hline
    {\bf SP$_2$} $({\bf q}_{\theta}, {\bf q}_{\phi})=(0,\frac{\pi}{2},\pi),(0,\frac{\pi}{2},0)$ &
    $(0,\frac{\pi}{2},\pi),(\pi,\frac{\pi}{2},0),(0,\frac{3\pi}{2},\pi),(\pi,\frac{3\pi}{2},0)$
    & $(0,0,\pi),(\pi,0,0)+(\pi,\pi,0)+(0,\pi,\pi)$
    \\\hline
    {\bf SP$_3$} $({\bf q}_{\theta}, {\bf q}_{\phi})=(0,\frac{\pi}{2},\pi),(\frac{\pi}{2},0,\frac{\pi}{2})$ &
    $(0,\frac{\pi}{2},\pi),(\pi,\frac{\pi}{2},0),(0,\frac{3\pi}{2},\pi),(\pi,\frac{3\pi}{2},0)$
    &
    $(\frac{\pi}{2},\frac{\pi}{2},\frac{3\pi}{2}),(\frac{3\pi}{2},\frac{3\pi}{2},\frac{\pi}{2}),
    (\frac{\pi}{2},\frac{3\pi}{2},\frac{3\pi}{2}),(\frac{3\pi}{2},\frac{\pi}{2},\frac{\pi}{2})$
    \\\hline
  \end{tabular}
  \caption{\label{tblsq}Candidate phases, the associated ${\bf q}_{\theta}, {\bf q}_{\phi}$,
    for the spirals, and the peak locations in the structure factor $S({\bf q})$. All the
    {\bf q} components have the same saturation value, given by $\frac{1}{2N_p}$, where $N_p$
    is the number of non-zero {\bf q} peaks in the S({\bf q}). $N_p=2$ for FM, A and C, $N_p=4$
    for $\up\up\dwn\dwn$ and SP$_1$,$N_p=6$ for flux and $N_p=8$ for SP$_2$ and SP$_3$. The factor
    of $\frac{1}{2}$ comes as we have half the spins at zero value, which halves the normalization.}
\end{table*}

\begin{figure}[b]
  \psfig{figure=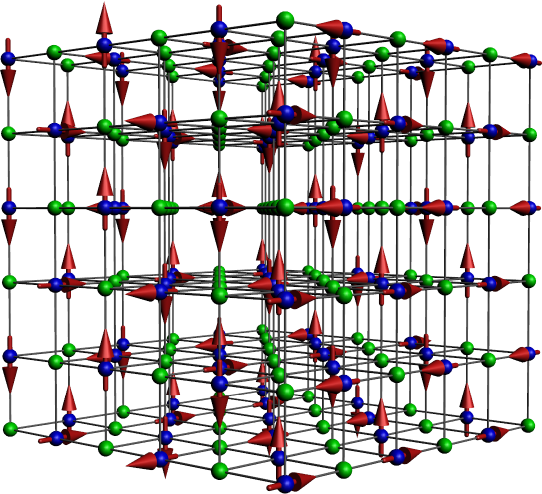,width=6.0cm,height=6.0cm,angle=0}
  \vspace{.2cm}
  \psfig{figure=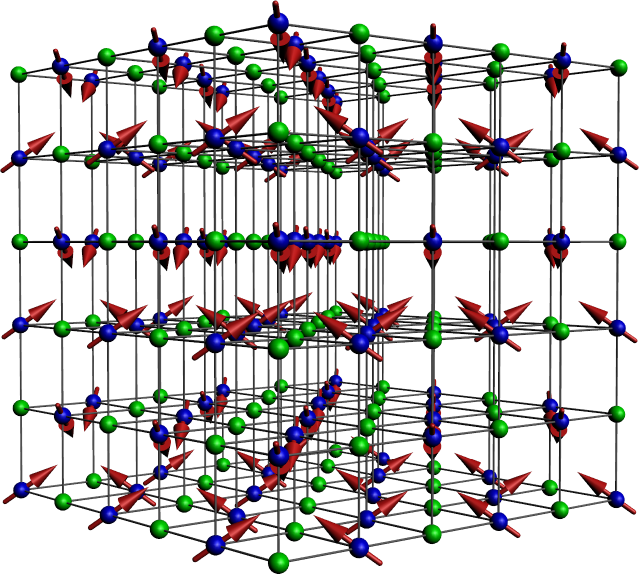,width=6.0cm,height=6.0cm,angle=0}
  \caption{ \label{spins_flsp}Colour online.
    Spin configuration for a typical spiral phase (top) and the
    `flux' phase (bottom). Since the spin configurations are non-collinear
    the electrons delocalise over the whole system. }
\end{figure}

\subsection{Monte Carlo method}
\label{mcmeth}
The model involves spins and fermions, and if the spins
are `large', $2S \gg 1$, they can be approximated as
classical. This should be reasonable in materials
like SFMO where $S = 5/2$.
Even in the classical limit these spins are annealed variables
and their ground state or thermal fluctuations have to
be accessed via iterative diagonalisation of the electronic
Hamiltonian.
We use a `traveling cluster' Monte Carlo (MC)
method where the cost of a spin update is estimated via
a small cluster Hamiltonian instead of diagonalising the whole
system~\cite{sanjeevTCA}.

We typically use a $12 \times 12 \times 12 $
system with the energy cost of a move estimated via a $4^3$ cluster built around the
reference site.
We principally track the magnetic structure factor
$$
S({\bf q} )={1 \over N^2} \sum_{{\bf r}, {\bf r}'}
\langle {\bf S}_{\bf r}.{\bf S}_{\bf r'} \rangle
{e}^{i {\bf q}  \cdot ({\bf r} - {\bf r}')}
$$
where $\langle ... \rangle$ denote thermal average.

Althogh the magnetic lattice is FCC, the electrons delocalise
on the combined B-B$'$ system which is
a cubic lattice. Hence we define our wave-numbers ${\bf q}$
with respect to the full B-B$'$ lattice.
As a result even a simple state like the ferromagnet
corresponds to peaks at ${\bf q} =(0,0,0)$ and ${\bf q} =(\pi,\pi,\pi)$
and not just ${\bf q} =(0,0,0)$.
This is because the spin field is also
defined on B$'$ sites and it has to have zeros on these sites.

This complication, and the possibility of spiral phases, {\it etc},
mean that (i)~there are multiple ${\bf q}$ values which could be
significant at low temperature, and (ii)~the $S({\bf q})$ peaks 
could be small even in the ordered state.
Combined with the intrinsic noise in MC data (which is
enhanced due to a complex energy landscape, discussed later)
it is sometimes
difficult to identify complicated ordered phases.
Therefore, to complement the MC results we have also used
the following variational scheme.

\subsection{Variational scheme}
\label{varsch}
We explore a  set of magnetic states, comparing 
their energy to locate the minimum within that family for
a fixed set of electronic parameters. We use:
\begin{eqnarray}
  \label{eq:spfam}
  {\bf S}_{\bf r}  & =&
  p({\bf r}) \{  {\hat x} 
  \sin\theta({\bf r} ) \cos\phi({\bf r} )
  + {\hat y} \sin\theta({\bf r} ) \sin\phi( {\bf r} ) \cr
  &&~~~~~~~ + {\hat z} \cos\theta({\bf r} ) \}  \nonumber
\end{eqnarray}
where
$ \theta({\bf r} ) = {\bf q}_{\theta}.{\bf r} $ and
$ \phi( {\bf r} ) = {\bf q}_{\phi}.{\bf r} $
with
$p({\bf r} )= 1$ if ${\bf r}  \in B$ and
$p({\bf r} )= 0$ if ${\bf r}  \in B^{\prime}$.
${\hat x}$, {\it etc.}, are unit vectors in the
corresponding directions.

The vector field ${\bf S}_{\bf r}$ is 
characterized by the two  wave-vectors 
${\bf q}_{\theta}$ and $ {\bf q}_{\phi}$.
For a periodic configuration, these should be
${\bf q}_{\theta} = \frac{2\pi}{L}(q_1,q_2,q_3)$ and 
${\bf q}_{\phi} = \frac{2\pi}{L}(p_1,p_2,p_3)$,
where $q_i$'s and $p_i$'s are integers, each of which
take $L$ values in $\{0,1,2,3,...L-1\}$.
There are $\sim L^{6}$ ordered magnetic configurations
possible, within this family, on a simple cubic lattice 
of linear dimension $L$.

The use of symmetries, {\it e.g.},
permuting components of $q_{\theta}$, {\it etc.},
reduces the number of candidates somewhat, but they
still scale as $\sim L^{6}$. 
For a general combination of ${\bf q}_{\theta},
{\bf q}_{\phi}$ the eigenvalues of H cannot be analytically
obtained because of the non trivial mixing of electronic
momentum states.
We have to resort to a real space diagonalisation.
The Hamiltonian matrix size is $2 N$ ($=2L^{3}$)
and the diagonalization cost is $\sim N^3$.
So, a comparison of energies based on real space diagonalisation
costs $\sim N^5$, possible only for $L \le 8$.

We have adopted two strategies: (i)~we have pushed this 
`${\bf q}_{\theta}, {\bf q}_{\phi}$'
scheme to large sizes via a selection scheme described 
below, and (ii)~for a few collinear configurations,
where Fourier transformation leads to a small matrix,
we have compared energies on sizes $\sim 400^3$. 

First, scheme (i).
For $L=8$ we compare the energies of all possible
phases, to locate the optimal pair
$\{ {\bf q}_{\theta}, {\bf q}_{\phi} \}_{min}$ 
for each $\mu $. 
We then consider a larger system with a set of 
states in the neighbourhood of $\{ {\bf q}_{\theta}, {\bf q}_{\phi} \}_{min}$.
 If we consider $\pm \pi/L$ variation about
each component of ${\bf q}_{\theta,min}$, {\it etc}, that involves
$3^6$ states. The shortcoming of this method is that it explores
only a restricted neighbourhood, dictated by the small size result.
We have used $L = 12,16,20$ within this scheme.

The phases that emerge as a result of the above process
are (i) Ferromagnet (FM), (ii) A-type, (iii)  C-type,
(iv) `flux', and (v) three spirals SP$_1$,SP$_2$,SP$_3$.
A-type is consists of ($1,1,1$) FM planes 
with alternate planes having opposite spin orientation 
(see Fig.\ref{fig:spinA} top panel).
If we convert each of these planes to alternating FM lines, so that
the overall spin texture is alternating FM {\it lines} in all directions,
we get C-type phase (see Fig.\ref{fig:spinC}.

The `flux' phase is different from the spiral families described
using period vectors ${\bf q}_{\theta},{\bf q}_{\phi}$. It is the
augmented version of `flux' phase used in cubic lattice double exchange
model by Alonso {\it et al}\cite{PhysRevB.64.054408}(Table-I). It
has spin-ice like structure, and is described by
$$
{\bf S}({\bf r})=\frac{p({\bf r})}{\sqrt{3}}((-1)^{y+z},(-1)^{z+x},(-1)^{x+y})
$$
The spiral SP$_n$ phases are characterised by comensurate values of
${\bf q}_{\theta},{\bf q}_{\phi}$ (See Table~\ref{tblsq} for details of
periods and the S(q) peaks). 

The simplest, SP$_1$ can be viewed
as $\frac{\pi}{2}$-angle pitch in the $(110)$,$(101)$ and $(011)$
directions. The other two spirals SP$_2$ SP$_3$ are respectively
C-type and A-type modulations upon SP$_1$. Just as flipping
alternate $1,1,1$ planes in a FM leads to the A type phase,
flipping the spins in the $(111)$ planes 
alternatively in SP$_1$, leads to SP$_3$.
Analogously, flipping FM lines in a  FM and leads to 
C-type order- and a similar exercise on 
SP$_1$ leads to SP$_2$. This modulation is also seen in the
$S({\bf q})$ peaks of SP$_2$ and SP$_3$. See the Table~\ref{tblsq}, where
all the three spirals have 4 S({\bf q}) peaks common, and SP$_2$ and SP$_3$
possess extra S({\bf q}) peaks of the A-type and C-type correlations.

In scheme (ii)  we take collinear phases from the phase diagram
via Monte-Carlo and variational scheme~(i), and compare
them on very large lattices. 
This does not require real space diagonalisation.
The simple periodicity of these phases 
leads to coupling between only a few
$\vert {\bf k} \rangle $ states. The resulting small matrix
can be diagonalized for the eigenvalues and these summed numerically.
We also did it for the `flux' phase, where the resulting matrix is
a bit larger, but still it gets us access to eigenvalues for the `flux'
phase on large lattices. The details of this calculation are
discussed in Appendix~\ref{appdx}. Where the collinear phases
(and `flux') seem to dominate the phase diagram we compute phase
boundaries by calculating the energy on very large lattices.

\begin{figure}[t]
  \psfig{figure=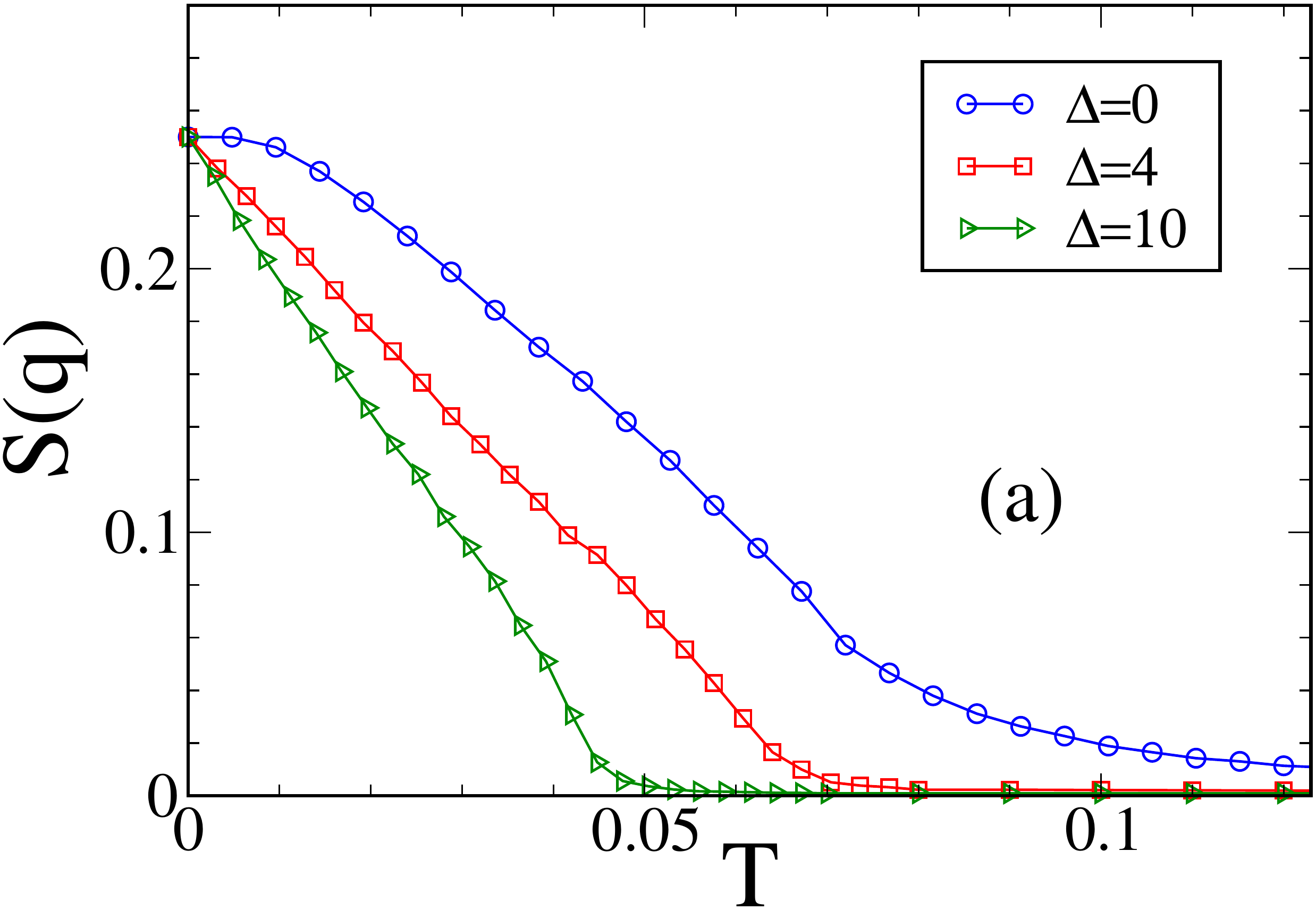,width=7.0cm,height=4.5cm,angle=0}
  \psfig{figure=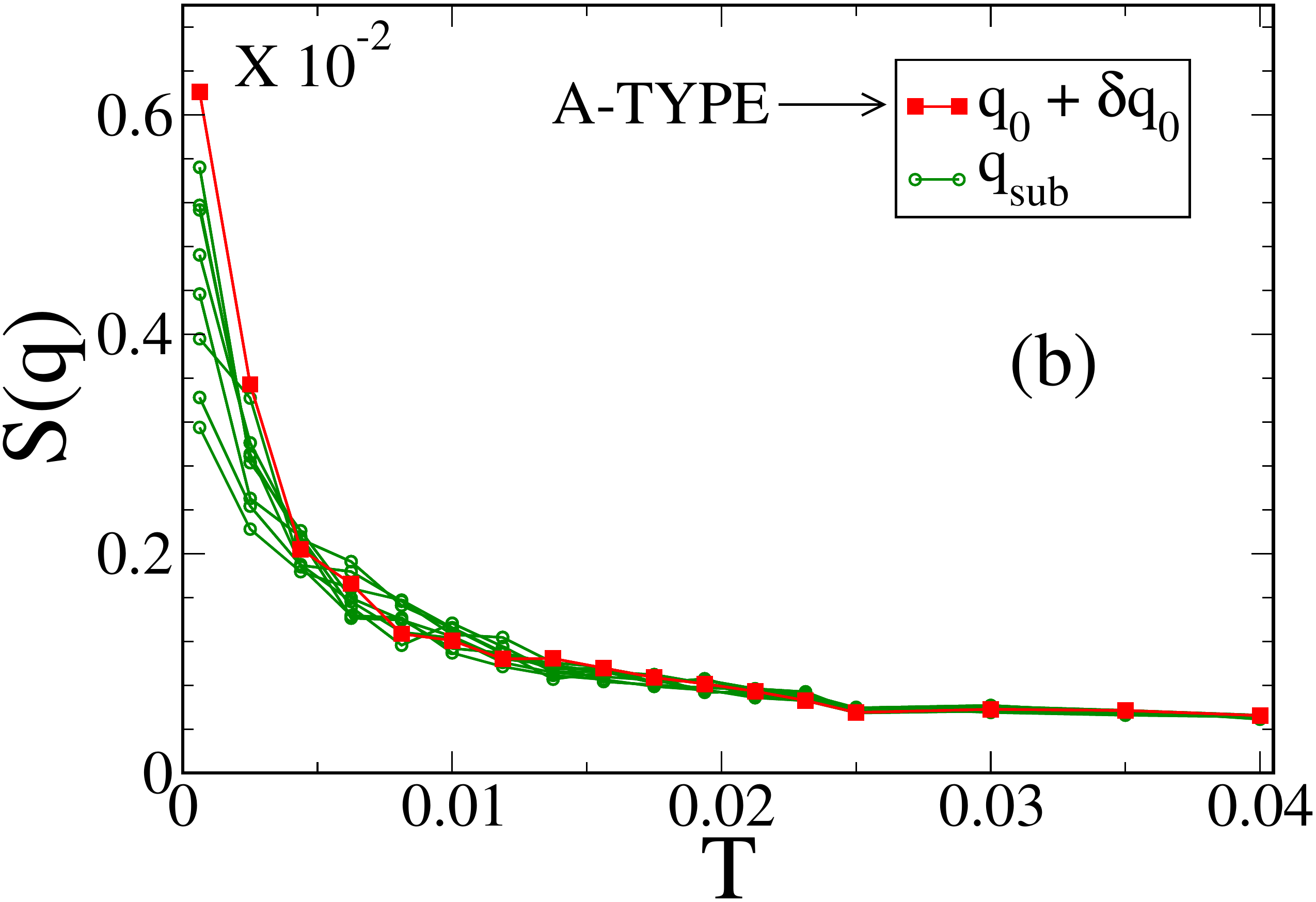,width=7.3cm,height=4.5cm,angle=0}
  \psfig{figure=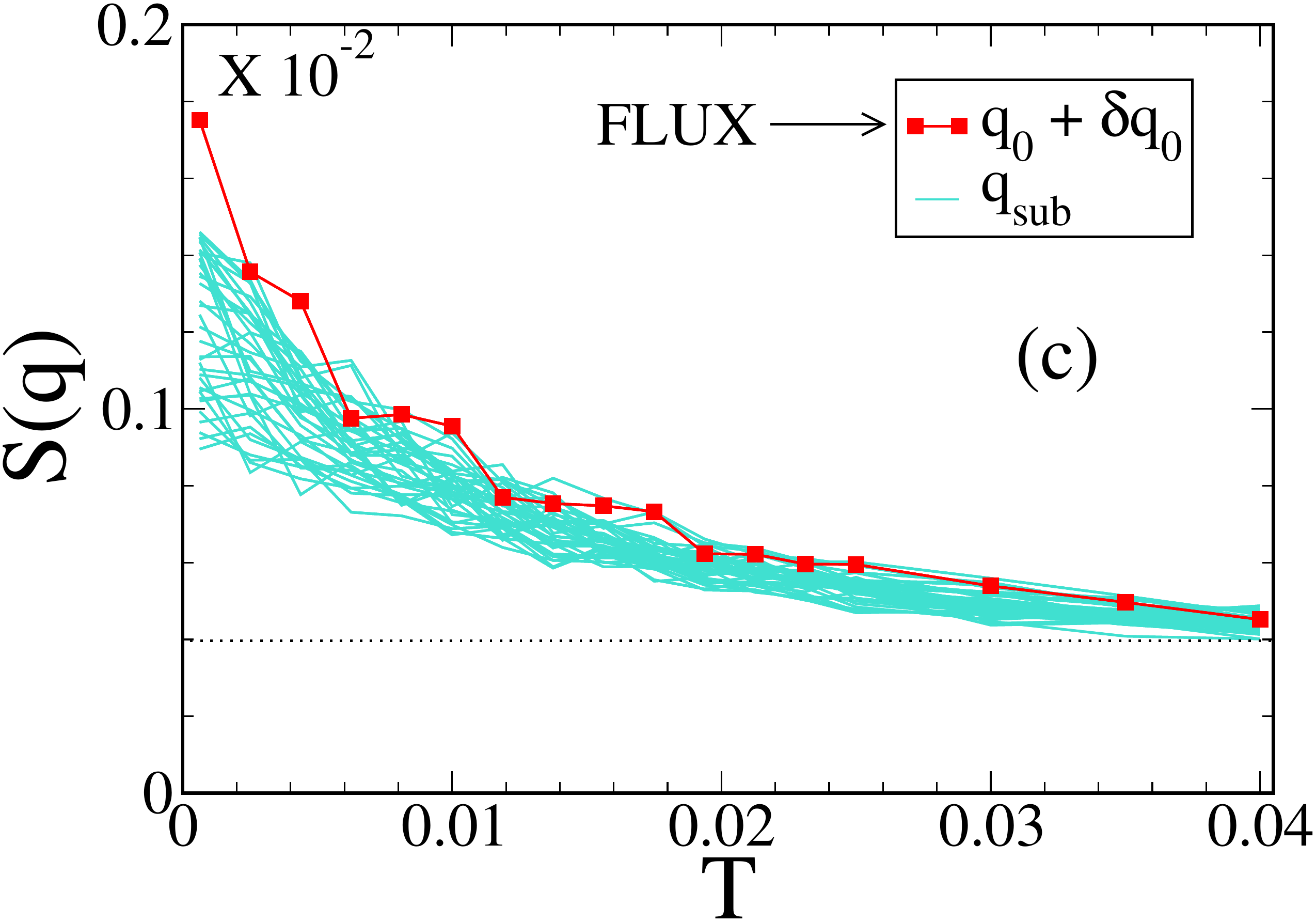,width=7.3cm,height=4.5cm,angle=0}
  \caption{\label{sqplt1}Colour online: Temperature dependence of
    structure factor peaks for three typical densities and $t'=0$.
(a).~For $\Delta=0,~4,~10$, ferromagnetic order at $n=0.20$.
    (b).~The growth of A type correlations (and the noise around the
    principal peak, at $n=0.50$.
    The ordering wave-vector
    $\roarrow {q_{0}}$ is listed in Table-\ref{tblsq}.
    $\delta \roarrow {q_{0}}$ are $\sim {\cal O}(\frac{1}{L})$
    (c).~`flux' type correlations at  $n = 1.50$. The features
    are at and around the ordering wave-vector in Table-\ref{tblsq}.
Note the scale factors on the $y$ axis in (b) and (c).
  }
\end{figure}

\section{Particle-hole symmetric case}
\label{ph-sym-case}

The electrons move on the cubic lattice divided into two FCC 
sub-lattices each of which accommodate B and B$'$ sites. For
each of these sub-lattice, one can define particle-hole
transformation~\cite{fazekas} for B and B$'$ sub-lattices as
$f_i\rightarrow f^{\dagger}_i$ and $m_{i\sigma}
\rightarrow-m^{\dagger}_{i\sigma}$.
This transforms the Hamiltonian as 
$H_{particle}(\D,t,t')-\mu N \rightarrow H_{hole}(-\D,t,-t')-(\mu-\D)N$.
When $t'=0$, this simplifies to $H(\D,t)-\mu N \rightarrow H(-\D,t)-(\mu-\D)N$
which reflects in the phase diagram as the repetition of the phases
after half-filling. Introducing the $t'$ hopping destroys this symmetry,
but a reduced symmetry still remains relating $(\D,t,t')\rightarrow(-\D,t,-t')$,
which is reflected in the phase diagrams of particle-hole asymmetric case.

We first discuss the case of particle-hole symmetry, {\it i.e},
$t'=0$, and the case of $t' \neq 0$ in the next section. 
For each of these 
cases we first discuss the MC results since these are
unbiased, though affected by finite size and the cluster
update mechanism. This provides a feel for the relevant 
candidate states that we can explore more carefully within
a variational scheme. It also provides an estimate of $T_c$,
not readily available within the variational scheme.

\begin{figure}[t]
  \vspace{.01cm}
  \psfig{figure=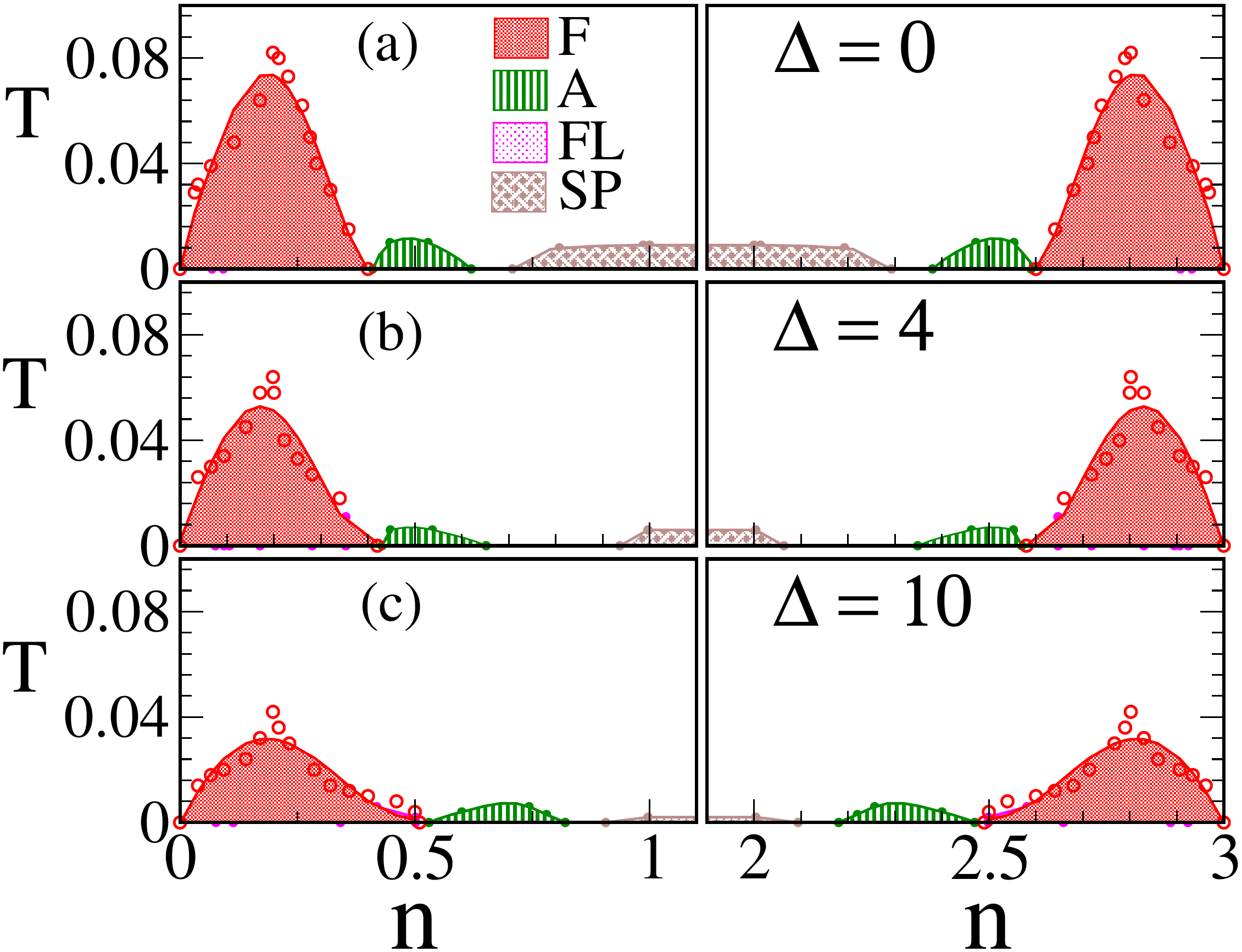,width=8.0cm,height=9.0cm,angle=0}
  \vspace{.01cm}
  \caption{\label{tc3del}
    Colour online: $n-T_c$ diagram for $\D = 0,4,10$ (top to
    bottom rows) as estimated from the Monte Carlo. Starting
    from low density ($n=0$) towards high density ($n=1$), we
    find FM with high T$_c$, thin window of A-type with very
    low T$_c$ as compared to FM, followed by `flux' in
    $\Delta=0$ and `spiral' in larger $\Delta$ case. The
    symbols are the actual MC estimated T$_c$, while the
    smooth lines are fit to the data.
  }
\end{figure}

Following this we show the ground states and phase separation (PS) windows 
that emerge from the variational calculation for varying $n$ and $\Delta/t$.
We also provide an alternate estimate of the ``$T_c$'' of these phases by
calculating the energy difference $ \delta E(n) =(E_{pm}(n) - E_{ord}(n))/N_{s}$, 
that the system gains via magnetic ordering. Here $E_{pm}$ is the electronic
energy averaged over disordered
(paramagnetic)  spin configurations while $E_{ord}$ is the
energy of the magnetically ordered ground state, both at the same electron
density $n$. $N_s (=\frac{N}{2})$ is the number of spins
in the system.
The phases that dominate the phase diagram are listed in Table-\ref{tblsq},
with the associated ${\bf q}_{\theta},{\bf q}_{\phi}$, and
the peak locations in the structure factor.

\begin{figure*}
  \begin{center}
    \psfig{figure=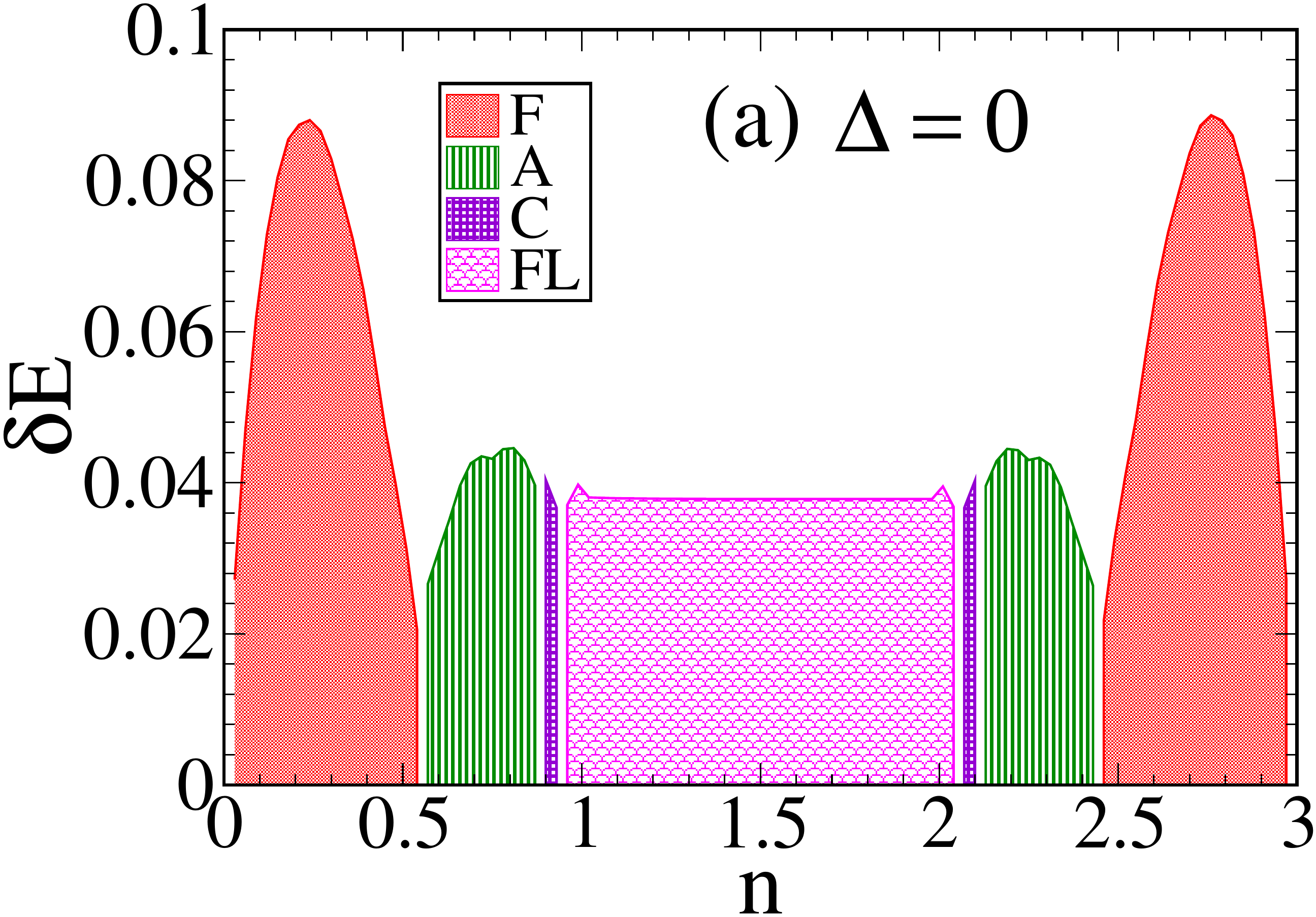,width=5.5cm,height=4.5cm,angle=0}
    \psfig{figure=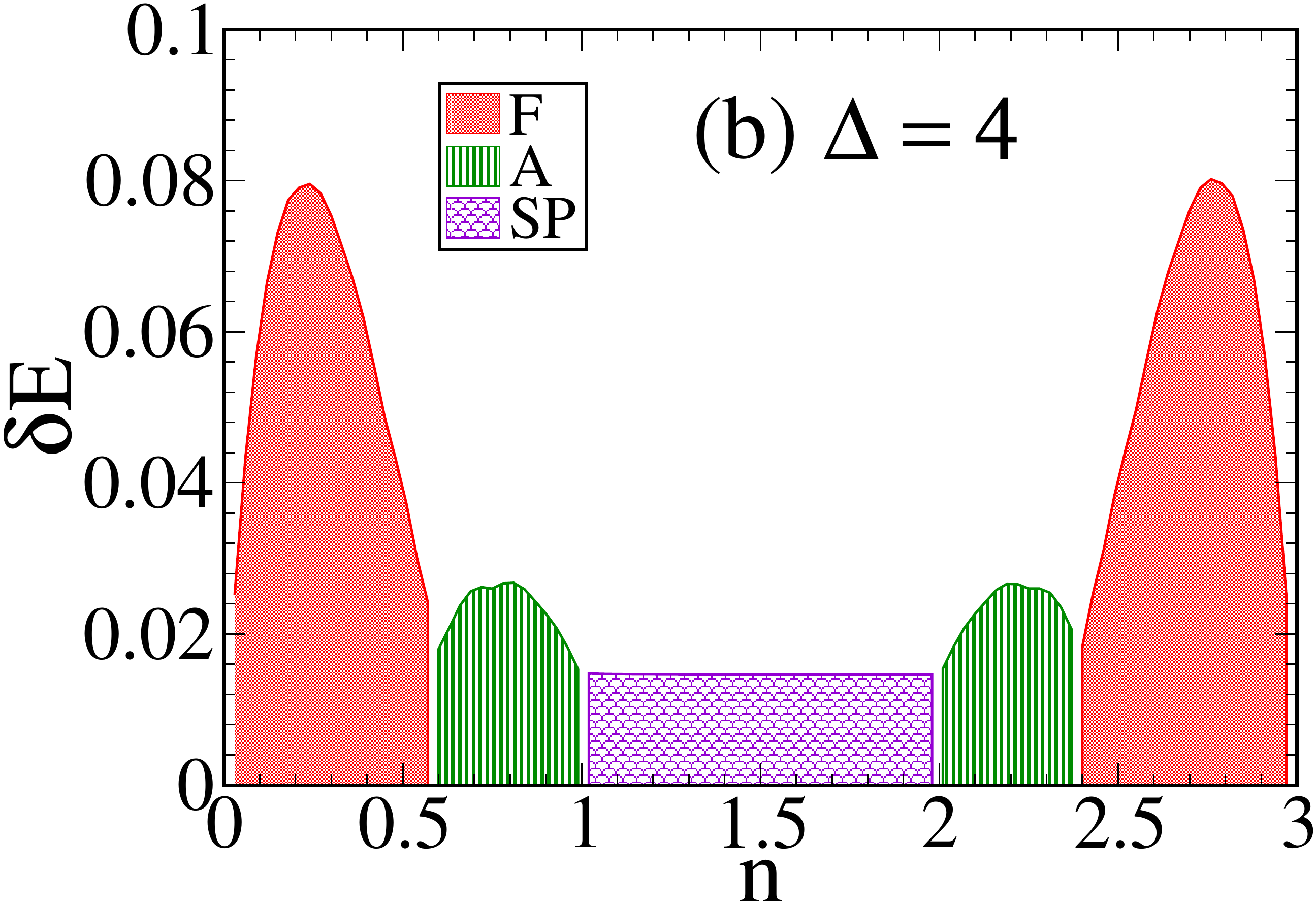,width=5.5cm,height=4.5cm,angle=0}
    \psfig{figure=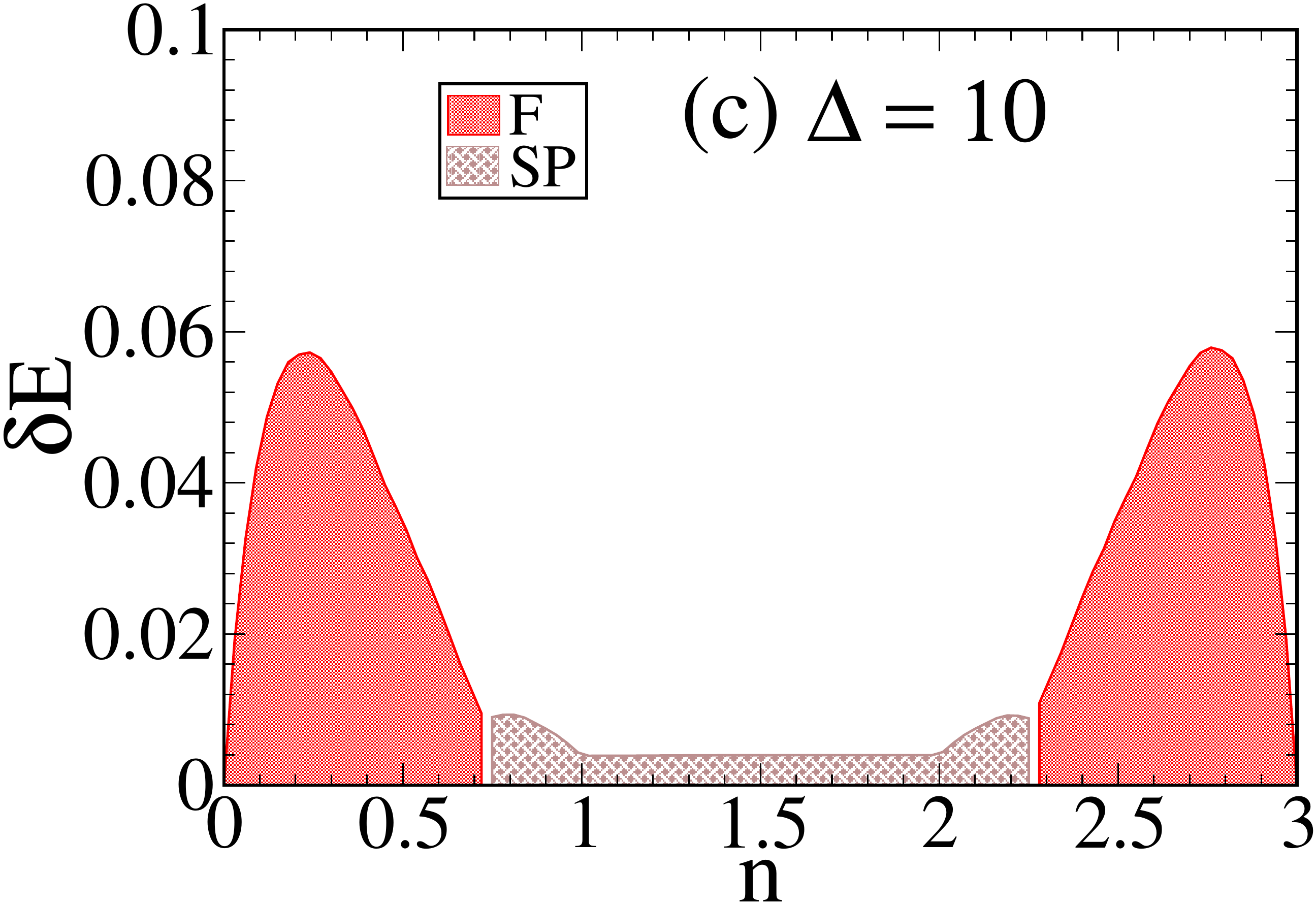,width=5.5cm,height=4.5cm,angle=0}
  \end{center}
  \caption{\label{delE}
    Colour online: The energy difference
    $\delta E$ of ground-state and paramagnetic phase. A variational
    estimate of the $T_c$ for three values of $(a)\D = 0,(b)\D = 4,
    (c)\D = 10$ and $t'=0$. The sequence of phases from low density
    to middle is FM, A, C and `flux' ($\Delta = 0$) or spiral
    ($\Delta = 4,10$). The decrease in the `T$_c$' with $\Delta$
    is more drastic in AF phases.
  }
\end{figure*}

\subsection{Monte Carlo}

We studied a  $N=12^3$ system using the cluster based update scheme. 
We used a large but finite $J$ to avoid explicitly projecting out
any electronic states~\cite{footnote1},
since that complicates the Hamiltonian matrix but allows only
a small increase in system size.
The magnetic phases were explored for  $\D = 0,~4$ and $10$.
An illustrative plot of peak features in $S({\bf q})$  as function
of temperature $T$, is shown in Fig.\ref{sqplt1} for some typical
densities, where, for FM,  S({\bf q}$_{FM}$)
shows monotonic decrease of T$_c$ with increasing $\Delta$. 
For A-type and `flux' phase,
the S({\bf q}) data shows a number of sub-dominant {\bf q} peaks
whose number keeps increasing as we move to more
complicated phases with increasing density.

Using the structure factor data, we establish the $n-T_c$
phase diagram for $\D=0,4,10$ that is plotted in Fig.\ref{tc3del}.
The Monte Carlo captures mainly three collinear
phases, namely FM, A-type, and a $\up\up\dwn\dwn$ phase.
The $\up\up\dwn\dwn$ phase corresponds to two FM
up planes followed by two FM down planes and so forth.
As the carrier density is increased via increasing 
$\mu$, we find a FM phase followed by the  A-type AF. A
$\up\up\dwn\dwn$  phase appears in a thin window surrounded
by FM itself. We suspected that this as a finite size effect,
and a comparison with the energy of the FM on larger lattices
($20^3$), shows that the FM is indeed the ground state in 
the thermodynamic limit, and so we consider FM and $\up\up\dwn\dwn$
collectively as FM only, and presence of $\up\up\dwn\dwn$ is not
indicated in the phase diagram.

The FM is stable at the ends of the density window, and its 
region of occurence is slowly
enhanced as we increase $\D$, see Fig.2 as well. 
The $T_c$ however decreases 
with increasing $\D$ since the degree of B-B$^{\prime}$ mixing (and
kinetic energy) decreases.

With further increase in $n$ the 2D system is known to make a 
transition to a line-like phase, and then a `G type' phase
(up spin surrounded by down, etc). In 3D  one would expect the
FM to change to a `planar' (A type) phase, then a `line like'
(C type) phase and finally to a G type phase if possible. All
of these are of course collinear phases, and geometric
constraints may lead to non-collinear order as well.

While we do access the A type phase with some difficulty,
our Monte Carlo cannot access the long range ordered C type
phase. However, we see clear evidence of C type correlations
in the structure factor. Comparing the energy of the ideal C
type phase with the short range correlated phase that emerges
from the MC we infer that such order is indeed preferred.
However, we cannot estimate a reliable $T_c$ scale.
In the next section we will see that the variational calculation
confirms the stability of the C type, among collinear phases,
in this density window, and will get a rough estimate of the
$T_c$ from the energy $\delta E$.

The G type phase is geometrically disallowed on the  B sub-lattice
due to its FCC structure. An examination of the structure factor
in the density window $n =[1,2]$ suggests  `flux' like correlations
at small $\Delta$ which evolves into a spiral at larger $\Delta$.
The frustration reduces the $T_c$ of the phases in this density window
compared to that of the FM. We studied the situation in
2D, where the system is unfrustrated, and the numbers below highlight
the impact of frustration. In 2D, Monte Carlo 
results yield $T_c^{AF}/T_c^{FM} \lesssim 1$, 
while in 3D $T_c^{AF}/T_c^{FM} \lesssim 0.1$.
We had focused on AF states at $n \sim 1.5$.
 If we compare the
$(\delta E)^{AF}/(\delta E)^{FM}$ for 2D and 3D, the numbers come
out to be $\sim 1.1$ and $0.5$ respectively~\cite{footnote2}. 
The comparisons
suggest a significant decrease in the binding energy (and hence $T_c$)
of the AF phases relative to the FM as we move from 2D to 3D.

\begin{figure}[b]
  \psfig{figure=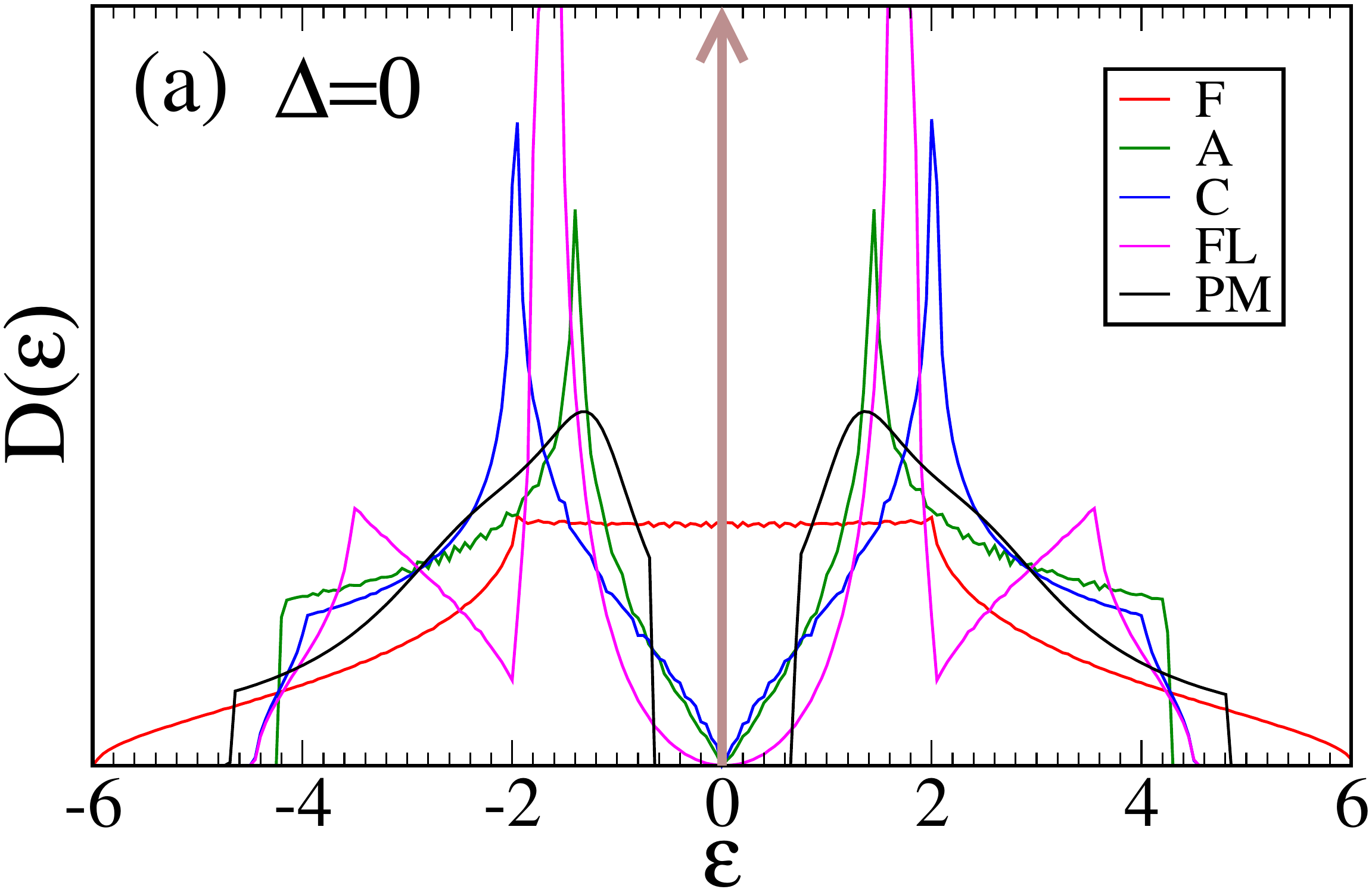,width=7.cm,height=5.cm,angle=0}
  \psfig{figure=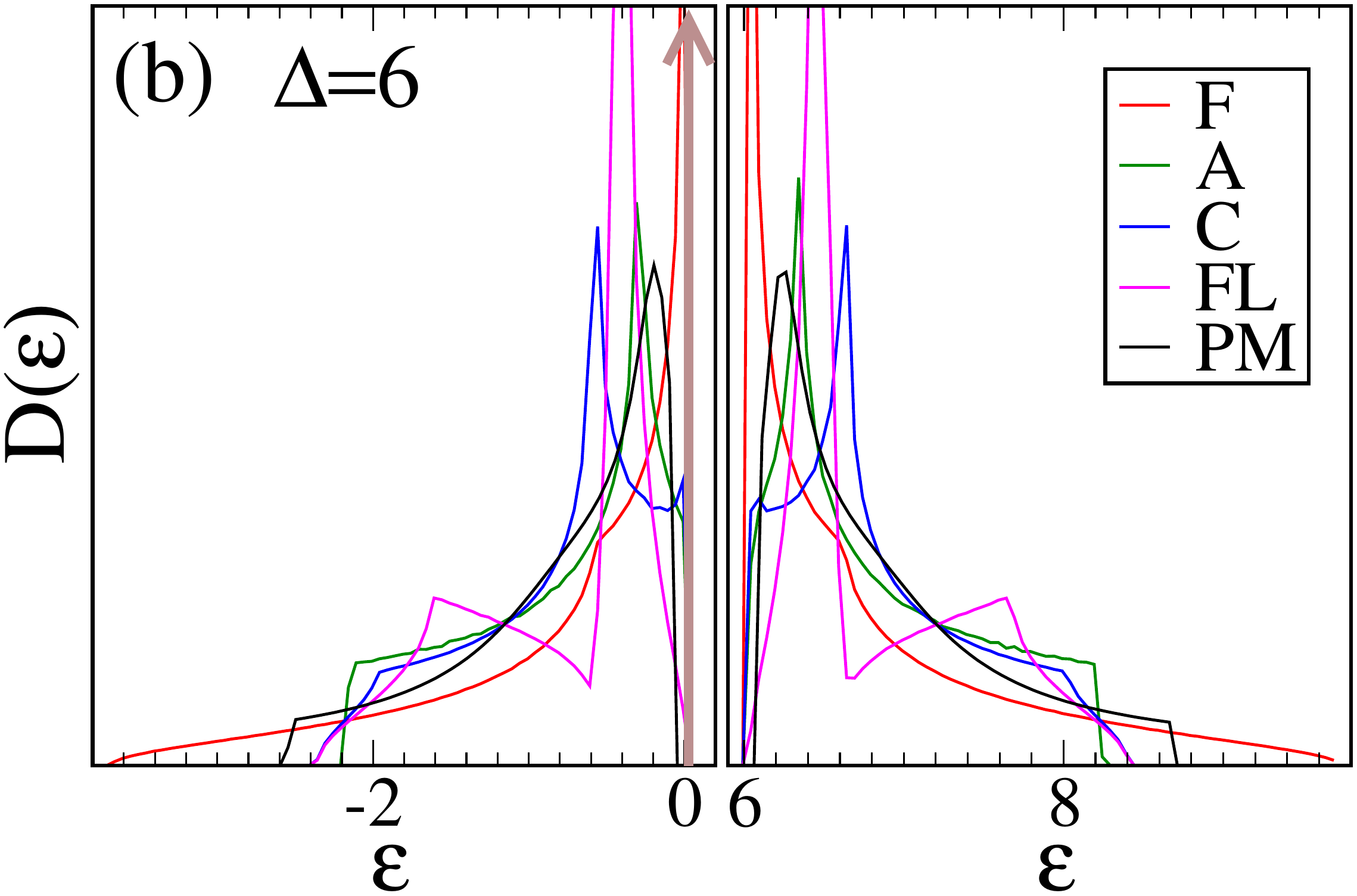,width=7.cm,height=5.cm,angle=0}
  \caption{\label{dos1}
    Colour online:
    density of states for the F,A,C,FL (`flux') and PM (paramagnetic)
    phases. (a) for $\D=0$ and (b) for $\D=6$. In order of decreasing
    band width are phase F,PM,C,FL, and A (FL and C have same bandwidth).
    This is for both $\Delta = 0,6$, and the order in general does not
    depend on the $\Delta$}
\end{figure}

When $t'=0$, the electron delocalisation happens through 
B-B$'$-B paths only 
(see the conduction paths, for example of collinear phases A and C in 
Fig.\ref{fig:spinA}~and~\ref{fig:spinC} respectively). In this cas all the phases
have an atomic level 
located at $\eps_{B'}(=0)$ in the limit $J\rightarrow \infty$.
This is directly seen in the density of states (DOS) of these phase.
In Fig.\ref{dos1} we show the DOS for the F, A, C, `flux' and
paramagnet phases. 
This dispersion-less level gives constant $T_c$ 
in density region $n=[1,2]$. This feature,
and several others, are modified
by finite B$'$B$'$
hopping, which leads to  broadening of this 
level.
It makes the DOS of the various magnetic phases asymmetric
(in energy) and also destroys the particle-hole symmetry in the phase diagram.
%Calculating the n-T$_c$ phase diagram for non-zero t$'$ (=0.3, See Fig.\ref{ntc-asym})

\subsection{Variational scheme}

Using the approach discussed earlier, we found the 
ground
state configurations at different electron
densities. In this set we also get certain spiral phases,
which are small variations of FM, A, and C phases in the left and
right part of the density window. Since in these parts Monte Carlo
also gives (for FM, A) clean result, we interpret it as a finite size
effect. To get convinced about it, we compare the energies of these
collinear phases (FM, A, C) with all their neighboring modulations
$\delta \roarrow q$s, at various lattice sizes. We find that with
increasing lattice size, the per particle energy difference between
collinear phase, and lowest energy candidate with the neighboring
${\bf q}_{\theta}$,${\bf q}_{\phi}$, decreases, which convinces us that
if we go to large enough lattice size, this difference will eventually
vanish and the collinear phases (F,A,C) will be the relevant candidates.

We use a similar scheme for the middle part, however there no simple
phase is suggested by this variational scheme (neither by Monte Carlo).
The phases we propose for the middle density part based on this
variational scheme are SP$_1$, SP$_2$, SP$_3$, and `flux'. 
See the configurations
in Fig.\ref{spins_flsp} and $S({\bf q} )$ details from Table~\ref{tblsq}.

In Fig.\ref{phdg1}, the  magnetic ground state 
is shown for $t'=0$ and $t'=0.3$ (top and bottom).
We see that for
$t'=0$ the phase diagram is symmetric in density.
 For small $\D$, in the range $0-4$, we have FM, followed by A-type,
C-type,
and `flux'  phase. The order reverses as we go in the other half of
the density window. The G-type phase which was largest stable phase in
2D (Fig.2 and Fig.5 in Ref.~\cite{sanyal:054411}) is almost taken over by
the `flux' phase. The stability of the `flux' phase 
decreases with $\D$ and it does not show up for $\Delta >4$.

In Fig.\ref{dos1}, we show the DOS for F, A, C, `flux' and PM phases.
The upper and the lower panel correspond to $\D = 0$ and $\D = 6$
respectively.  
In all the phases, at all $\D$, there is a spike (delta function)
at $\eps=0$, which accounts for the non-dispersive level at
$\eps_{B'}=0$. 

For FM all the core spins are $\up$ (say), so only $\dwn$ spin
electrons from B$'$ site get to delocalise while $\up$ spin
electrons remain localised at $\eps=0$, which corresponds to
the localised band in the sepctrum and the spike in the DOS.
So the localised level in FM is an $\up$~spin level. The nature
of this localised level however changes when we go to AF phases.
In collinear phases, its easy to understand the nature of this
localised band. Take for example the case of A-type in Fig.\ref{fig:spinA}
down panel with conduction path. The lattice
is divided into two sublattices (each of which are of layered
zigzag shape), blue and red, such that, if (say) core spins in
the $f$ sites in the blue sub-lattice are all $\up$, then the
same in red sub-lattice are $\dwn$. As a result, in the blue
sub-lattice, $\dwn$ spin electrons get to delocalise, while
$\up$ spin electron remain localised. the opposite happens
in the red sub-lattice. Since these lattices are disconnected
from each other, one can separately diagonalise them. But
each of this sub-lattice, however complicated in shape, is
a FM, so it gives $\frac{1}{3}$ of the levels localised at
$\eps=0$, which will be $\up$ spin in blue sub-lattice, while
it will be $\dwn$ spin in red sublattice. Since both the
sub-lattices have same number of sites/unit cells, we get
$\frac{1}{3}$ of the levels localised at $\eps=0$ but
now spin degenerate. The delocalised states have also
to be spin-degenerate, and their nature depends on the
way the conduction paths divide the lattice into two
sub-lattices. 

For each spin channel the conduction paths are
layered zigzag, 2 dimensional in the A type phase, while they
are 3 dimensional in the C type phase. 
%This which results in C type
%having the larger bandwidth than A type. 

This appearance of the localised band is not restricted to
just the collinear phases, but also happens 
for non-collinear
phases, and even the paramagent. 

\section{Particle-hole asymmetry}
\label{ph-asym-case}

The model with only `nearest neighbour' (BB$'$) hopping has a 
rich
phase diagram. However, this has the artificial feature of a
non dispersive level. In reality all materials have some degree
of B$'$B$'$ hopping and we wish to illustrate the qualitative 
difference that results from this hopping. We explored two
cases,  $t'=0.3$ and $t'=-0.3$
 for these particle-hole asymmetric cases.

\subsection{Monte Carlo}

\begin{figure}[b]
\hspace{-20pt}
\psfig{figure=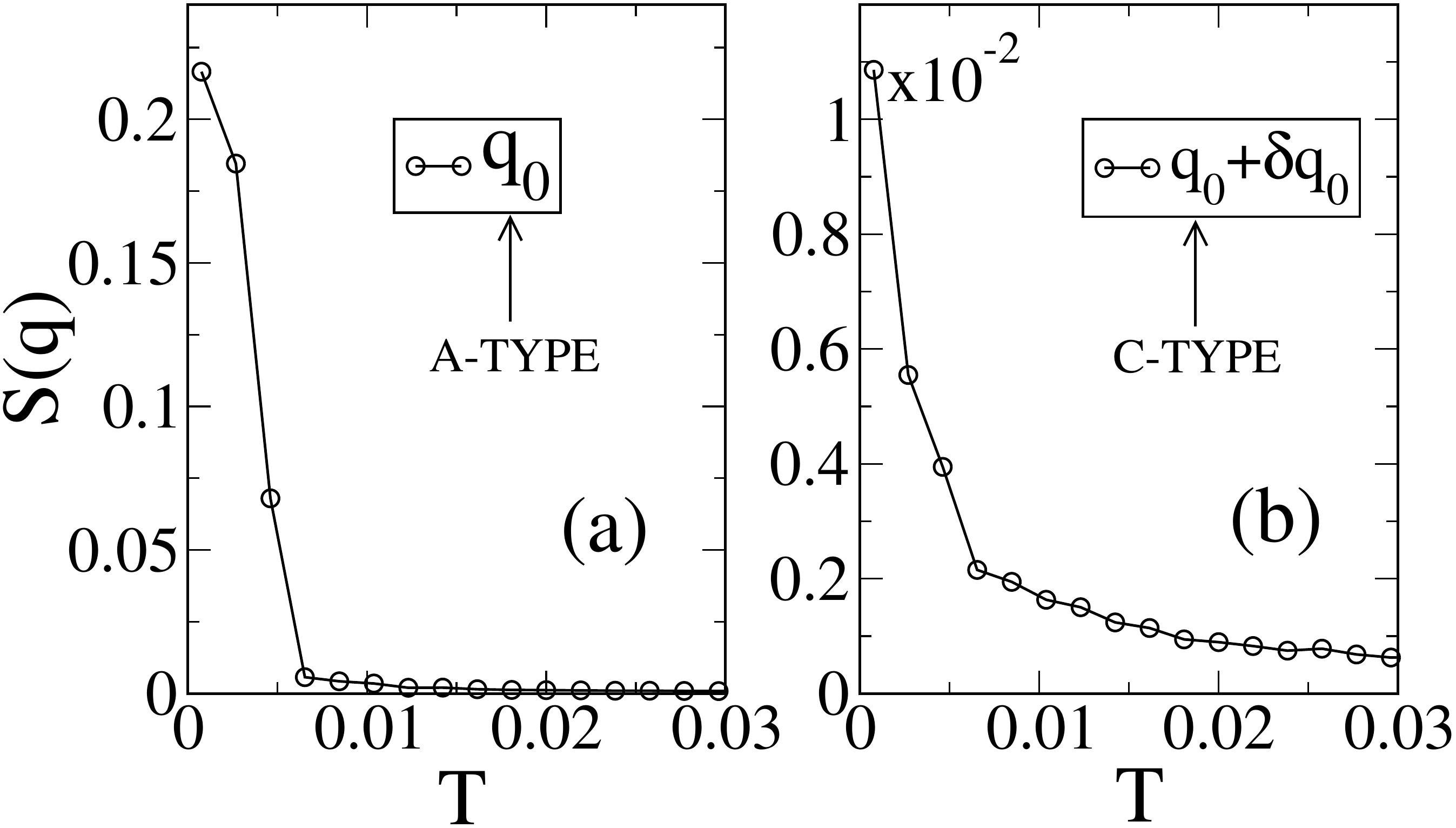,width=9cm,height=4.5cm,angle=0}
\caption{\label{sqplt2}
    $S({\bf q})$ for $t'=0.3$ and $\Delta=0$
    at (a): a typical density $n\sim 0.5$ for A-type phase
    and (b): a typical density $n\sim 1.2$ (this particular choice
    is for minimum frustration in C-type) for C type phase.
    A demonstration of $S(q)$ with no sub-dominant peaks,
    unlike at $t'=0$. }
\end{figure}

% MC/VC asym
%
\begin{figure}[b]
\psfig{figure=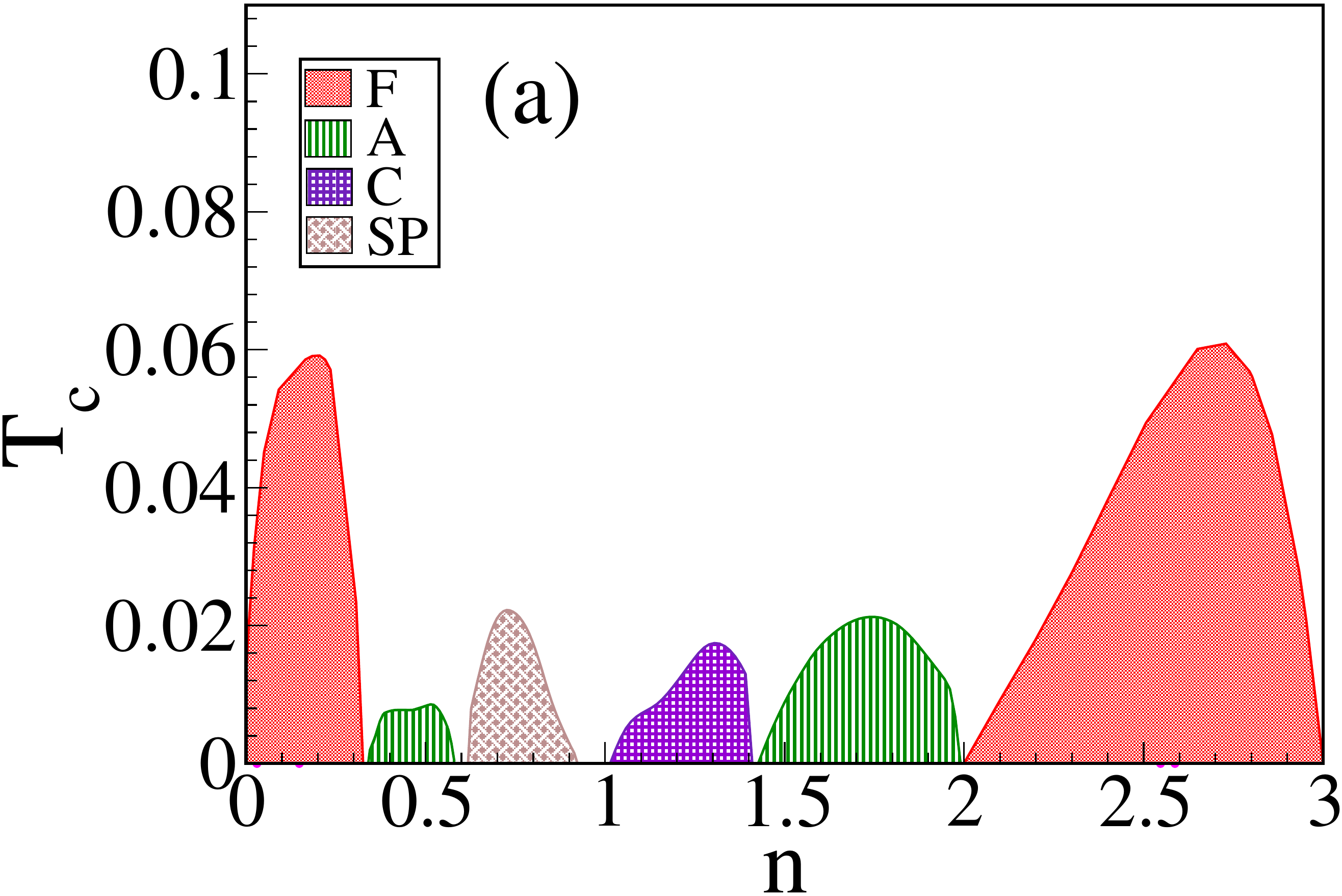,width=7cm,height=4.5cm,angle=0}
\psfig{figure=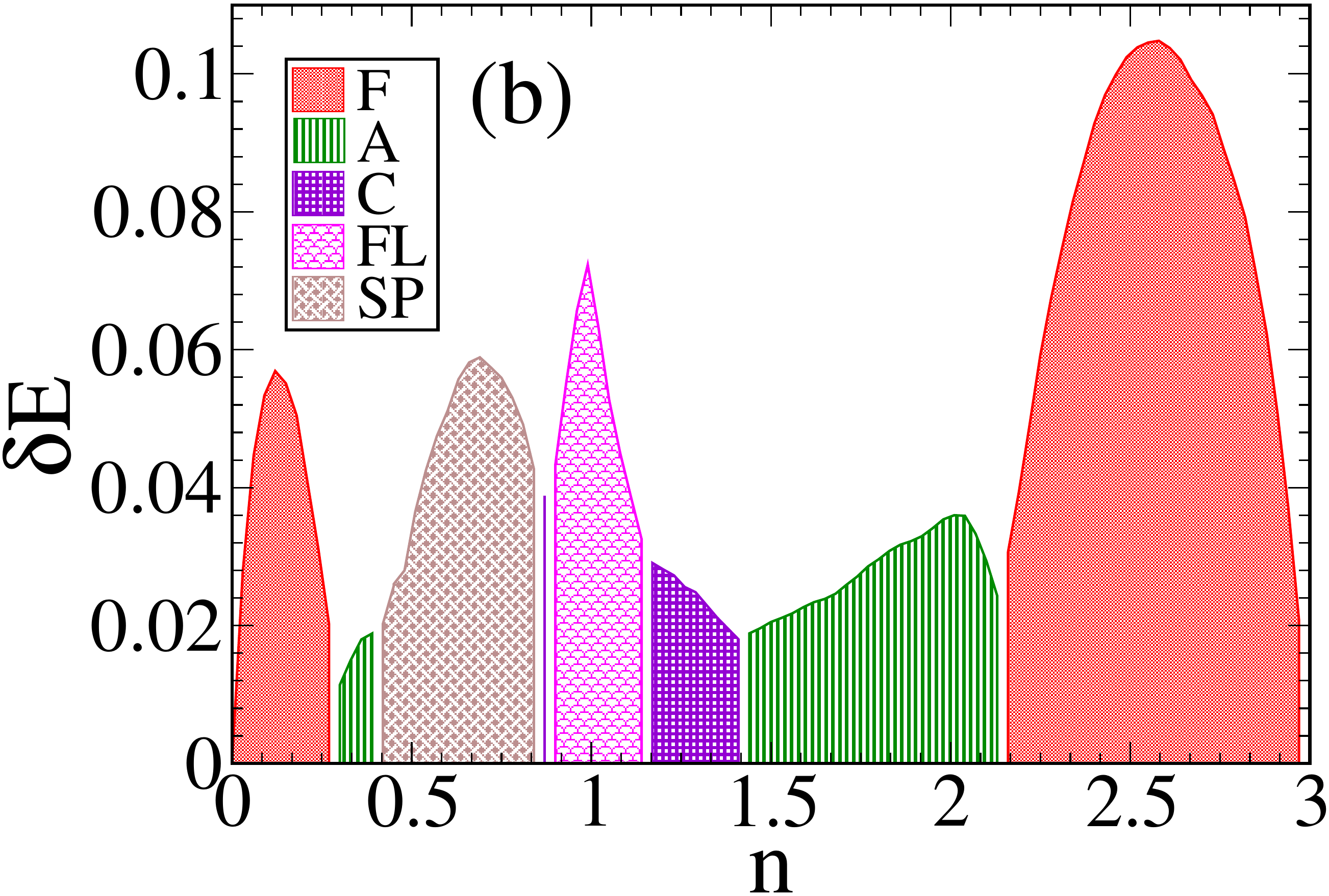,width=7cm,height=4.5cm,angle=0}
\caption{\label{ntc-asym}
Colour online: Phase diagram obtained via Monte Carlo (top) and from the
variational calculation (bottom) at $t'=0.3$ and $\Delta=0$.
}
\end{figure}

In Fig.\ref{sqplt2}(a),(b) we show the structure factor data,
at two densities, for (a) A type and (b) C type phases,
to demonstrate one remarkable difference from the particle-hole
symmetric case. As we saw earlier in Fig.\ref{sqplt1} for
$t'=0$ the structure factor data were
very noisy for AF phases, with many sub-dominant {\bf q} peaks
around the central peak.
  The saturation value for the A-type peak in the 
symmetric case was
$\sim 10^{-2}$, while now it is $\sim 0.2$, close to the
ideal value of  $0.25$.
The sharp change in the structure factor makes the 
identification of the $T_c$ scale more reliable.
Although inclusion of $t'$ does not remove
the noise completely, it is reduced over a 
reasonable part of the phase diagram.

Fig.\ref{ntc-asym}(a) presents
the $n-T_c$ phase diagram for $t'=0.3$ and $\D=0$
established from Monte Carlo, along with the $\delta E$
from the variational approach (Fig.\ref{ntc-asym}(b)).  
In this case, the phases that appear as a function of density $n$
are FM, A type, spiral, C type, A type and FM again.
For FM, the window of stability gets reduced in  the left (low density)
part but enhanced to almost full band ($n\sim 2$ to 3) 
in the right (high density)
part. The $\uparrow \uparrow \downarrow \downarrow$ 
phase appears again, but being a finite size artifact, is absorbed in
the FM (and not shown). The $T_c$ is usually reduced, from the symmetric
($t'=0$) case, as $BB'$ hopping provides conduction paths that are
non-magnetic. There is a wider space with moderate $T_c$ for
A type phase, located asymmetrically in density. Its more stable, in
the right window, than left window, hence possessing relatively higher $T_c$
than left. The correlations of spiral and C type phases are also 
captured with relatively less noise, see Fig.\ref{sqplt2}(b) for example
of C-type correlation. Although S({\bf q}) data for these phases still
contain some noise, so that we don't get clean ground state here either.
The $n-T$ phase diagram for
$t'=-0.3$ and $\D=0$, can be obtained from transformation
$n\longrightarrow 3-n$, {\it i.e.}, 
reversing the density axis of Fig.\ref{ntc-asym}(a),(b).

\subsection{Variational scheme}

%
% ph diag: asym
%
\begin{figure}
  \psfig{figure=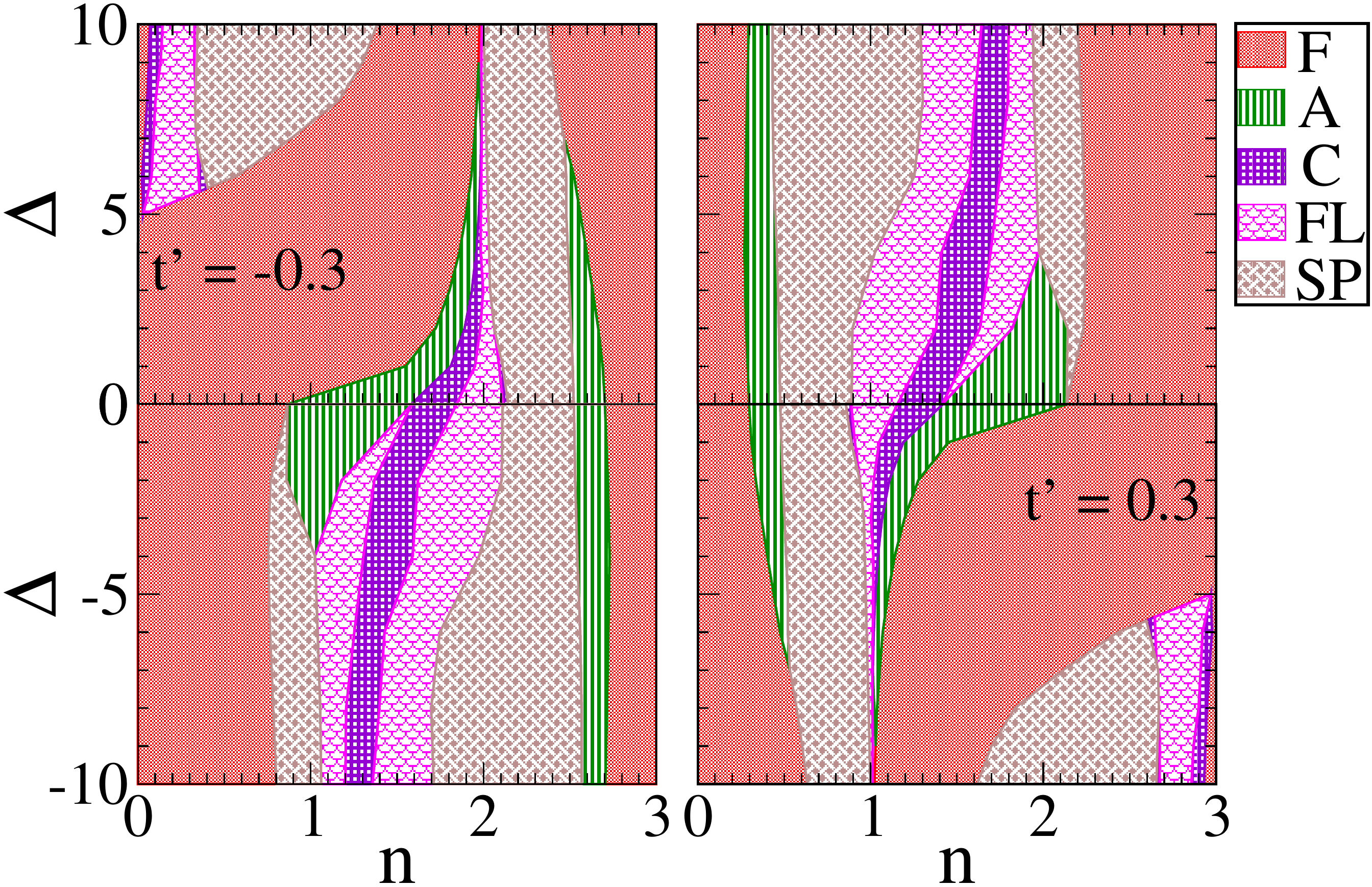,width=8.3cm,height=6.3cm,angle=0}
  \caption{\label{phdg2}Colour online: Ground state phase diagram 
in the presence of $t'$. Left panel: $t'$=-0.3. 
Right panel: $t'$=0.3.  } 
\end{figure}

\begin{figure}
  \psfig{figure=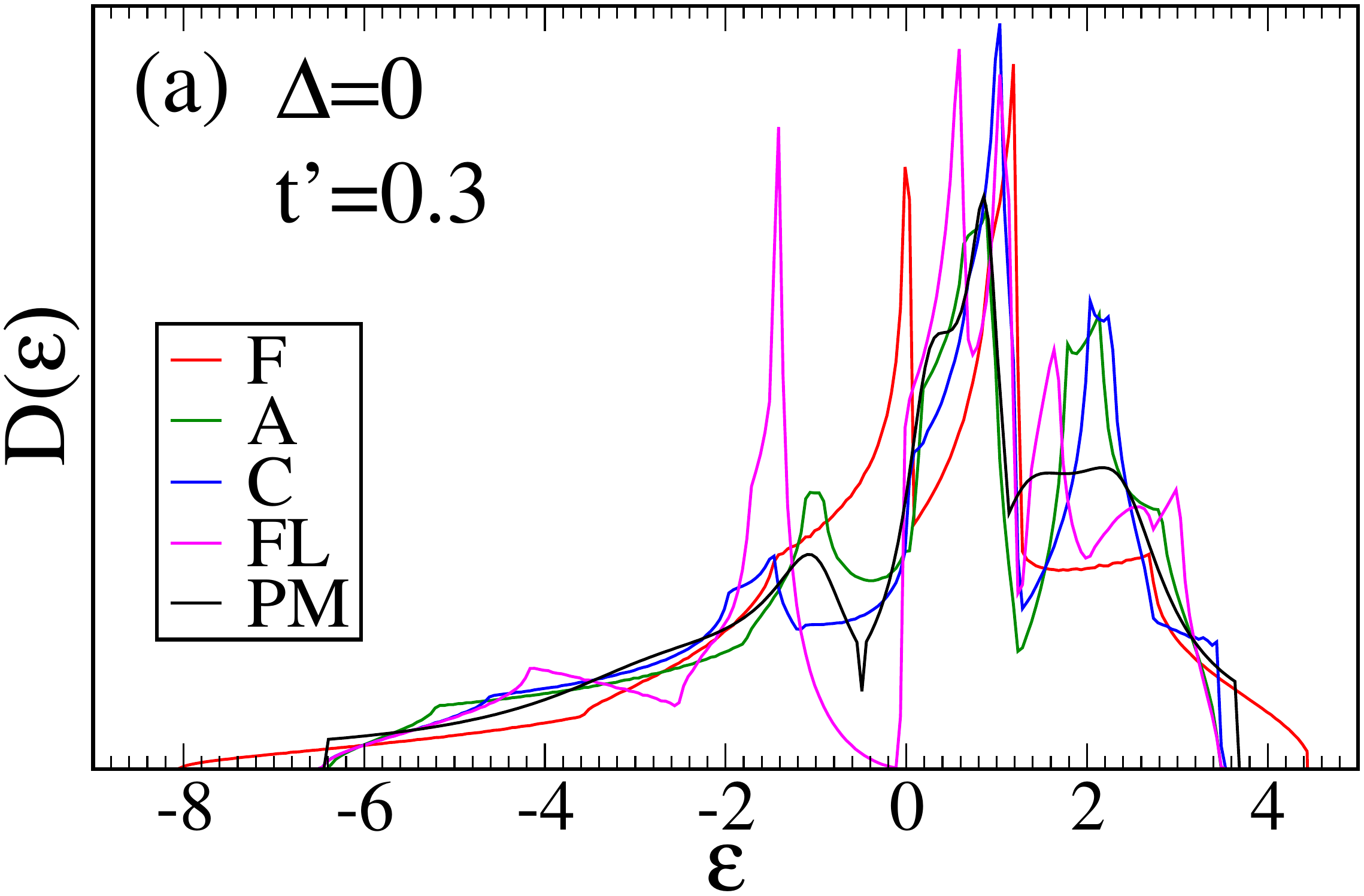,width=7.cm,height=4.6cm,angle=0}
  \psfig{figure=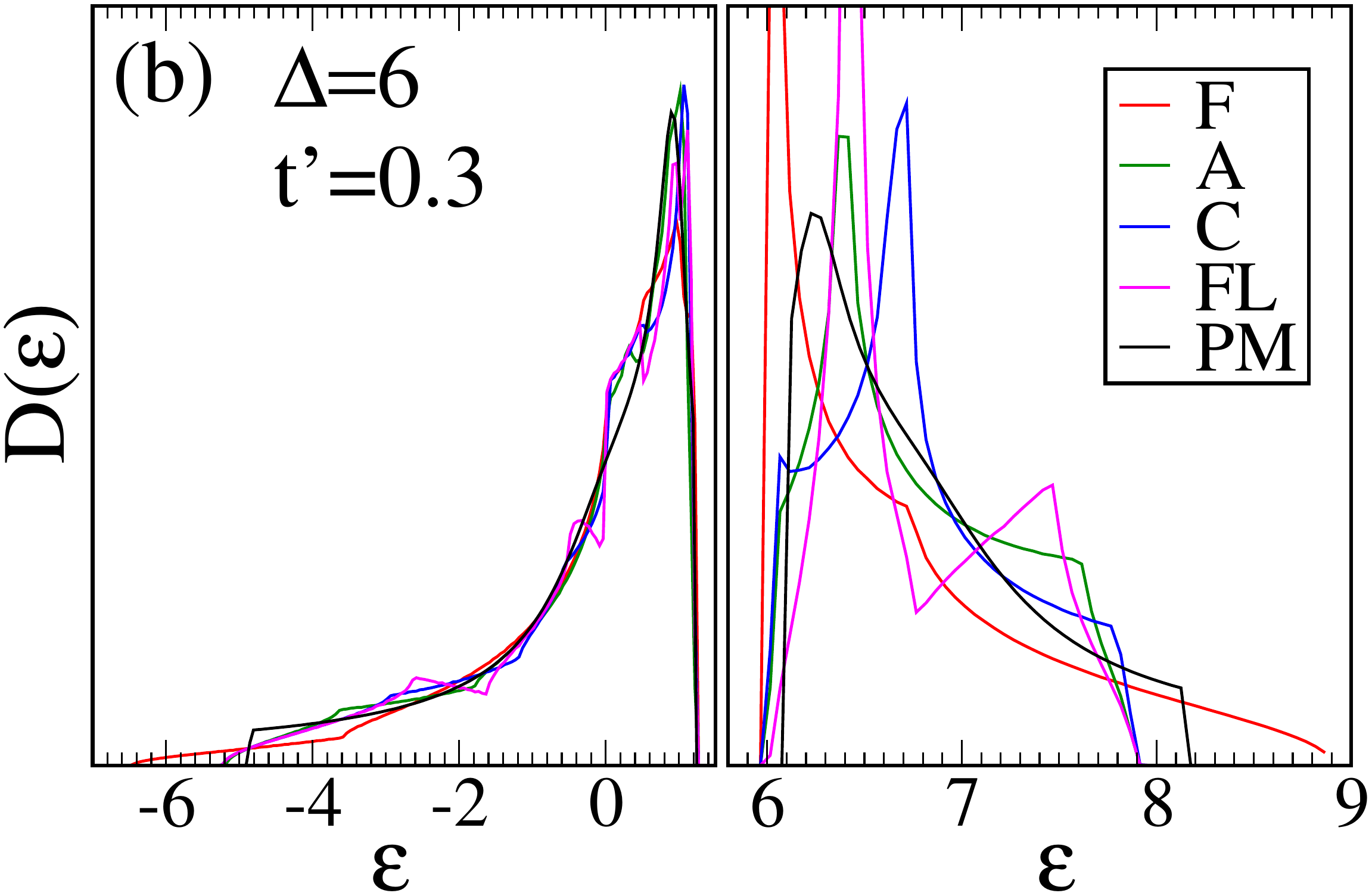,width=7.cm,height=4.6cm,angle=0}
  \psfig{figure=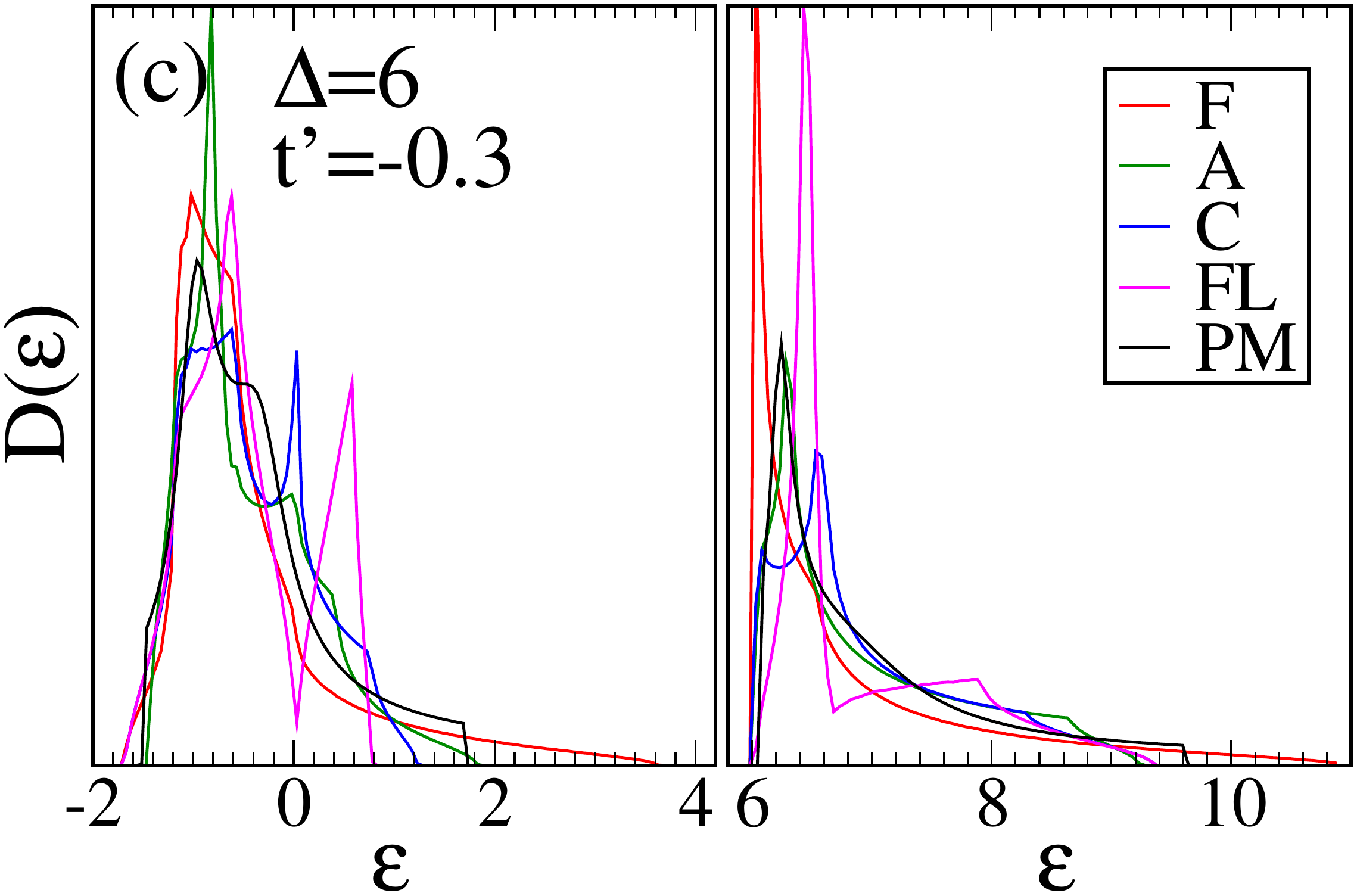,width=7.cm,height=4.6cm,angle=0}
  \caption{\label{dos2} Colour online:
    DOS for the F, A, C, FL, PM phases
    ~{(a):} $\D=0$, $t'=0.3$, there is no resemblence to the particle-hole
    symmetric case. FM and PM have largest bandwidth, while A, C, `flux'
    have almost same bandwidth
    ~{(b):} $\D=6$, $t'=0.3$, due to large band-gap, two bands
    are shown in two different panels (left, right) the same
    bandwidth order, right band less effected from $t'$, being
    situated arounf $\D$
    ~{(c):} $\D=6$, $t'=-0.3$. The structure of the band edges has
    changed drastically. Now band edges of the FM, `flux' and C type
    coincide on the left, while on the right edge of the first panal
    FM is more widespread.
  }
\end{figure}

We employ the variational scheme discussed earlier and obtain
the ground state phase diagram for $t'=\pm0.3$ is shown in
Fig.\ref{phdg2}. Turning on $t'$ has a significant effect
on the phase diagram, when we use the  $t'=0$ case, Fig.\ref{phdg1} top panel,
for reference. 
The particle hole symmetry $(n \rightarrow 3-n)$ is
destroyed at finite
$\Delta$ 
but a reduced symmetry $(n,~\D,~t')\rightarrow (3-n,-\D,-t')$
still holds. The phase diagram is richer in the middle of the
density window where crossing among various phases occurs at
different densities. Due to the symmetry 
mentioned above it is
enough to discuss the $\D > 0$ case with $t'=\pm 0.3$.

For $t'=0.3$ the trends from MC are well reproduced by the
variational scheme on large $(20^3)$ systems at $\D=0$.
We observe reduced stability of FM at low density 
and enhancement at high density. 

Note that the overall correspondence between the Monte Carlo 
and the variational approach is much better here than 
in the $t'=0$ case, Fig.~\ref{tc3del} and Fig.~\ref{delE}.

The A-type phase becomes very
thin in the left, but unaffected by $\D$, while in right side
it widens up in the low $\D$ and gets replaced by the spiral
quickly as we go up in $\D$. `flux' and C-type both become stable
for high $\D$ with a gradual shift in the high density.
For $t'=-0.3$, at very small $\D$ in the left and the middle
part A-type and the spiral are major candidates with small
window for C-AF and `flux'. The behaviour in this part is not
very sensitive to sign of $t'$. 

Focusing on $t'=-0.3$, 
as go up from $\D=0$ to $\D\sim 5$ the AF phases
become less and less stable and are almost wiped out from
the left part of the density, and FM becomes stable there. 
The largest stability window of FM occurs roughly near
$\D \sim 5$, where its stable upto $n \sim 1.8$. Going
further with higher $\D$, FM looses its stability, from C
type, `flux' and spirals. However, there is very thin strip
of stability of the FM in the band edge in the left part,
and towards the middle density, there is re-entrance of the
FM phase.

In the right part of the density, we have FM, A type and spiral.
Increasing $\D$ reduces the stability of A type to FM, making
it vanish near $\D\sim 7$, while FM window keeps increasing with
$\D$.

The DOS for the ordered F, A, C, `flux' phase and the paramagnet
are shown in Fig.\ref{dos2}. (a) For $t'=0.3$, $\D=0$ FM has the
largest bandwidth, with paramagnet second largest. The band edges of A, C,
`flux' almost coincide both for small and large $\Delta$. 
(b)~For $t'=0.3,\D=6$
the left band shows that all the phases seem to have `almost' similar
features in the DOS, while the right band shows distinct features of
each of the phases, similar to $t'=0$ DOS.
(c) For $t'=-0.3, \D=6$ however, has a disctint case. Here the lower edge of
the band for FM, `flux' and C coincide, and the DOS of `flux', or C, is
higher than FM, which explains why FM becomes unstable in the left
side and taken by `flux' and C, upon increasing $\D$.

\begin{figure}
  \psfig{figure=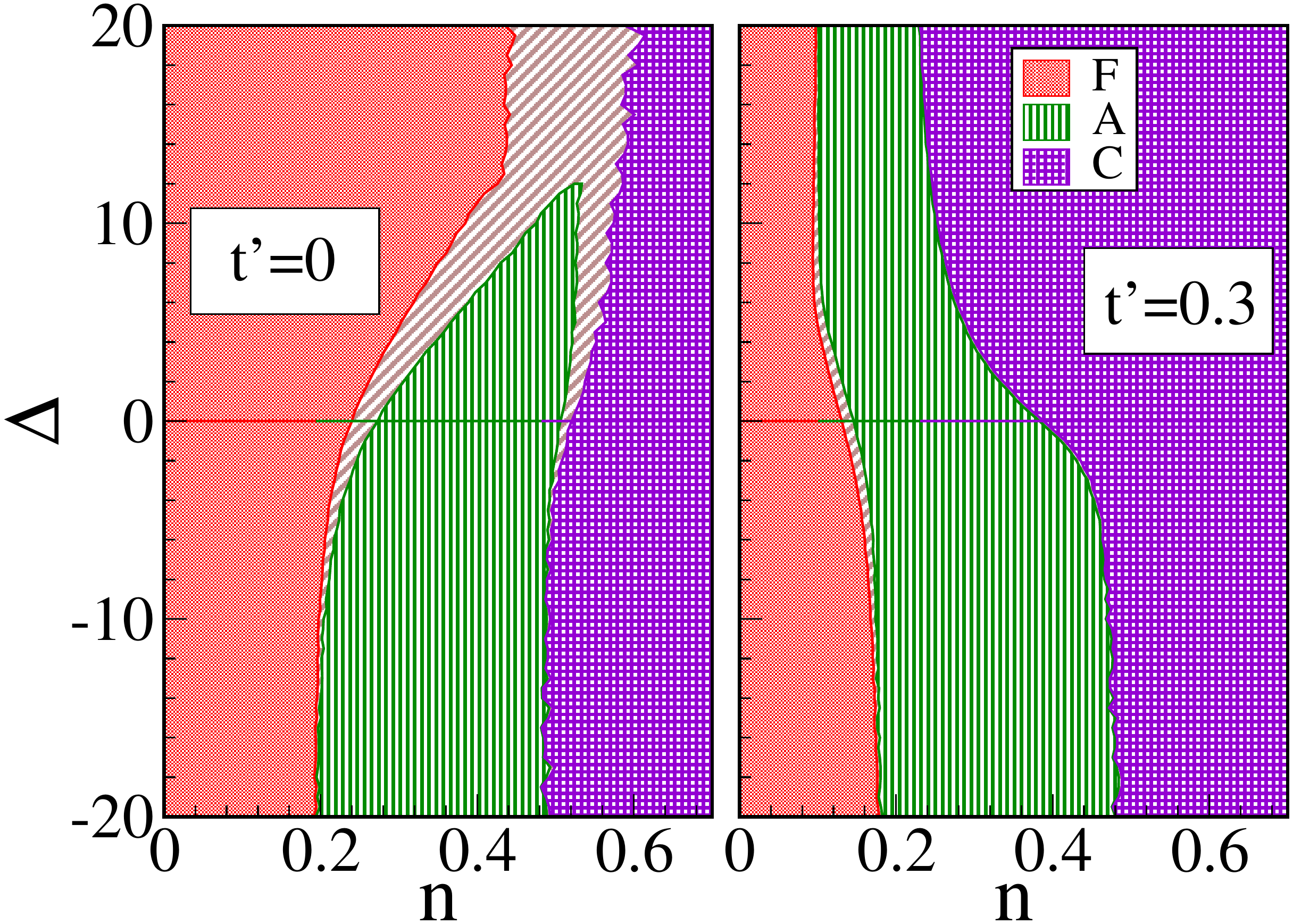,width=7cm,height=5cm,angle=0}
  \caption{\label{phsep}Colour online: PS regions. left column for
    t$'$=0 and right column of t$'$=0.3. Notice that PS regions are
    significant for $\Delta > 0$ and t$'=0$, for t$'\neq 0$ and for
    $\Delta < 0$ PS boundaries almost vanish.  }
\end{figure}

We also estimate the phase separation boundaries between FM, A, C phases
shown in Fig.\ref{phsep}. For $t'=0$ and for $\D>0$, we see that PS regions
are significant, while they vanish for $\D <0$ as we go down. For $t'=0.3$,
(right panel) the PS boundaries are too narrow to be visible.

%
% ph sepn, symm - asymm
% 

\begin{figure}[b]
  \begin{center}
    \psfig{figure=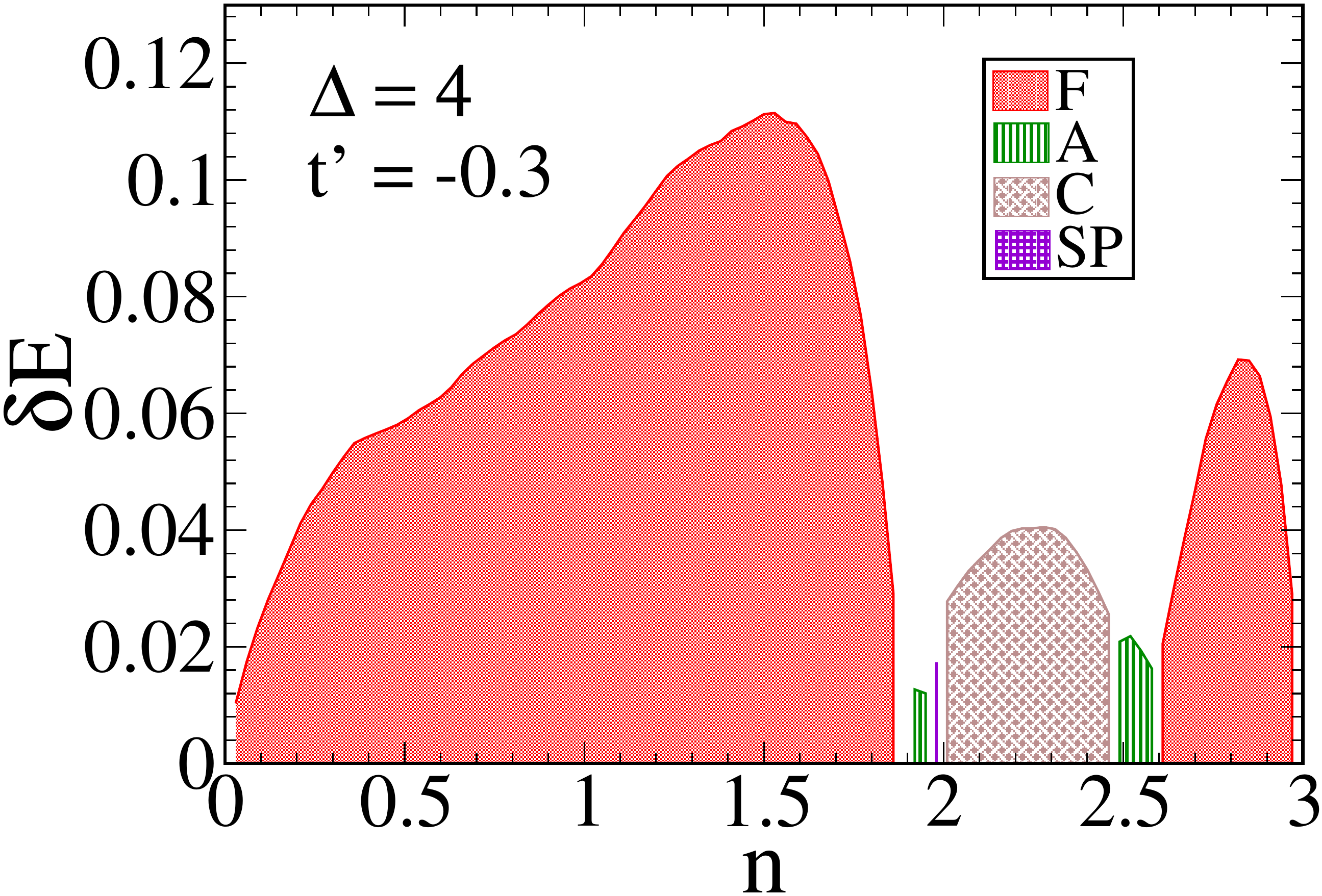,width=6cm,height=4cm,angle=0}
  \end{center}
  \caption{\label{delta-e} Colour online: Asymmetric case, the energy difference
    $\delta E$ of ground-state and paramagnetic phase.
    Top: $\D=4,t'=-0.3$ where FM is stable in the large portion of the density.
    Bottom: $\D=0,t'=0.3$ The trends of $\delta E$ match with T$_c$.
  }
\end{figure}

In Fig.\ref{delta-e} we have shown the $\delta E(n)$
calculated for 20$^3$ size, for $\D=4,t'=-0.3$, with the large stability window
of ferromagnet(See Fig.\ref{ntc-asym}). Here, though the $\D$ and $t'$ are
non-zero, due to unusually large stability window, the $\delta E$(or $T_c$)
is large.

To summarise, from the MC and variational 
data we learn that, apart from asymmetry in the
phase diagram, collinear FM and A type phases become stable in wide density
window. Their $T_c$ however is slightly reduced than the symmetric case.
The S({\bf q}) data showing less noise for A, C type and spirals indicates
that the energy landscape become `smoother' by $t'$ so that annealing process
becomes easier to get to the ground state. The energy differences $\delta E$
as well as MC estimated $T_c$s show overall decrease with $t'$. This is
understable as, by introducing $t'$ we allow electrons to more on the
`non-magnetic' sub-lattice B$'$. Now the energy of any phase, depends on
the energy gain via the hopping process. From the nearest $f-m$ hopping,
this gain scales as $\frac{t^2}{\D}$ subject to spin configurations, while
the from the next nearest $m-m$ hopping, this gain simply scales as $t'$,
and doens't care upon spin configurations. So more we increase $t'$ and $\D$,
the more we are making the energy of the system insensitive to
spin-configurations. The asymptotic limit of this is $\frac{t^2}{\D}\rightarrow 0$
when every phase has same energy as paramagnet. That also explains why the phase
seperation windows become very small with inclusion of $t'$.

In the couple of paragraphs below we try to create an understanding
of how the phase diagram is affected by $t'$. There is'nt,
unfortunately, an understanding of the effects over the 
entire density window, but we can at least motivate the 
changes at low density.

For $t'=-0.3$, 
the FM loses its stability to AF phases even at low $n$.
That is puzzling since one would expect  the FM phase to
have the largest bandwidth. 
We recall that in the $t'=0$ case, there is a localised band
coming from B$'$ level for all the phases. The dispersion of this
previously localised level causes the $m$ and $f$ 
to have a 
$ {\bf k}$ dependent separation, which was $\Delta$
for all ${\bf k}$ in the  symmetric case. The separation for these
levels in
 the asymmetric case is $\D_{\bf k}=\Delta - \eps'_{\bf k}$, which varies
from $\Delta - 12|t'|$, to $\Delta + 12|t'|$ in 3D. In 2D it varies from
from $\Delta - 4|t'|$, to $\Delta + 4|t'|$.

If we consider the simpler 
2D case for illustration, 
$\eps'_{\bf k}=-4t'\cos{k_{1}}\cos{k_{2}}$,
which, for $t'>0$ is minimum at {\bf k}=$(0,0),(\pi,\pi)$ while
the  maximum is at
{\bf k}=$(\pi,0),(0,\pi)$. For $t'<0$ the opposite will happen. 
Thus, for $t'>0$
{\bf k}=$(0,0),(\pi,\pi)$ and neighbouring states will experience enhanced
mixing $\sim \D-4|t'|$, while, the states near {\bf k}=$(\pi,0),(0,\pi)$
experience lower mixing. In the ferromagnet (both in 2D and 3D), the lowest
eigenvalue corresponds to ${\bf k}=(0,0)$, while (for 2D) the
G-type phase
has lowest eigenvalues at ${\bf k}=(0,0),(\pi,0)(0,\pi)(\pi,\pi)$.
Therefore, for $t'<0$, the lowest 
eigenvalues of both the phases are enhanced but the
band-edge of FM stays lower than G type. While in the other, the strongest
mixing states are $(\pi,0)$ and $(0,\pi)$, which aren't at the edge for FM,
its band-edge gets lower enhancement, while the band-edge of G type gets
lowered. For a given $t'$, as we increase $\D$, a point comes where band edges
of the FM and G, coincide. This is the point where FM loses its stability.

\begin{figure}
  \begin{center}
    \psfig{figure=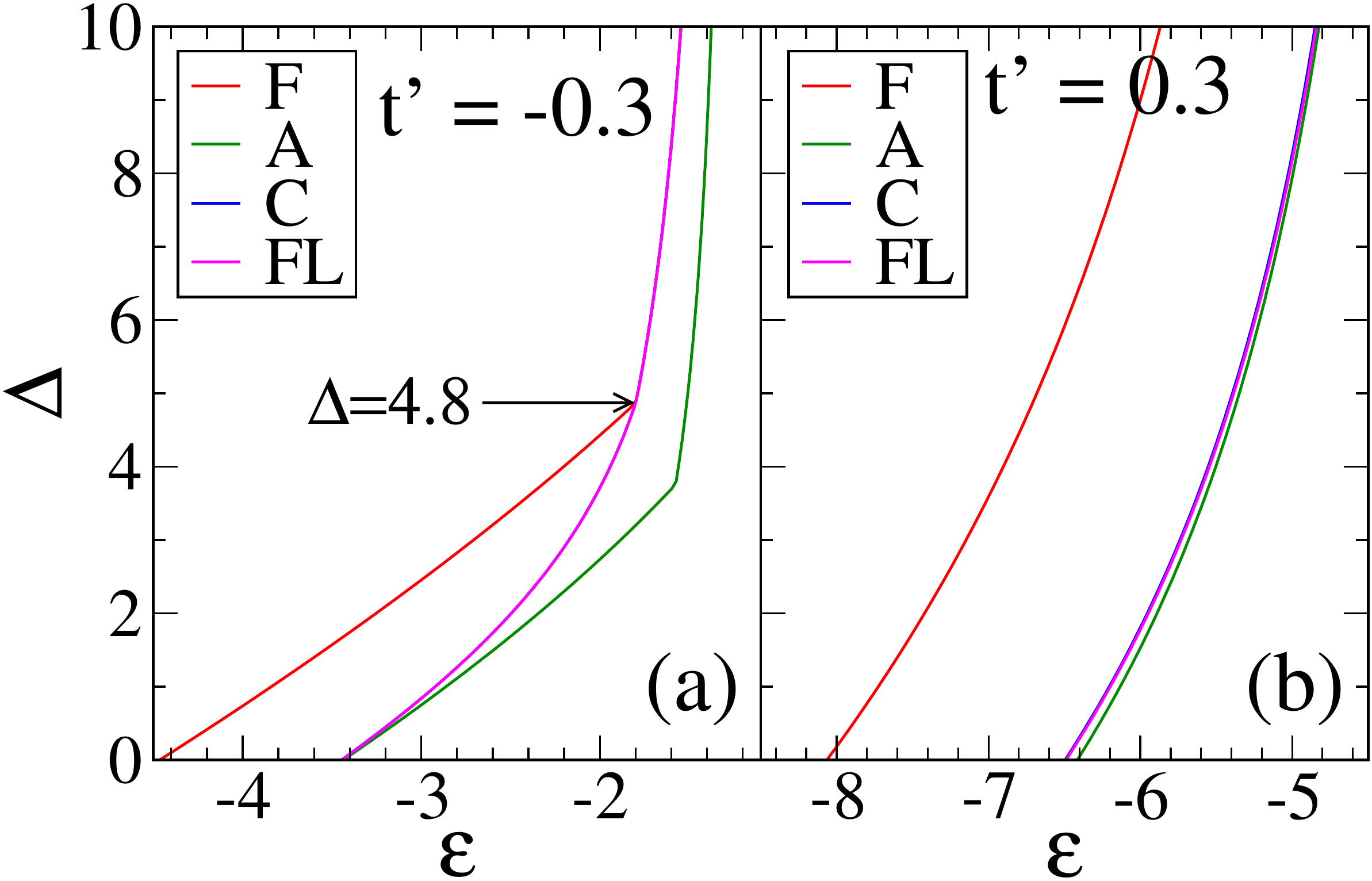,width=7cm,height=4cm,angle=0}
  \end{center}
  \caption{\label{edge} Colour online: Asymmetric case, lowest eigenvalues plotted as
    function of $\D$ for the F,A,C and flux phases calculated from the dispersions.
    (a) $t'=-0.3$ and (b) $t'=0.3$
  }
\end{figure}

The same arguement can be extended to 3D, with C and flux phases, just the
role of the ${q}$s gets extended to 3D (e.g, $(\pi,0,0)$ etc), and the 
correction in the separation is $\sim 12t'$ instead of $\sim 4t'$. 
In Fig.~\ref{edge}
we have shown the plot of lowest eigenvalues of F,A,C and flux phases with $\D$
for $t'=\pm0.3$

Finally, a comment (mainly a conjecture), Fig.18,
 on how the energy
landscape  of the DP model changes on addition of $t'$. We already 
know that the `binding energy' and $T_c$ of magnetic phases
reduce with increasing $t'$ - but also that the `noise' in the
cooling process also reduces quickly. 

If $t' \gg t$ then the electrons could delocalise
on the wide $t'$ based band populating the non magnetic
sites only. Magnetic order would make little  difference
to electronic energies and the `energy landscape' in the
space of spin configurations would be featureless, panel
(c) in Fig.18. There are no global minima, {\it i.e} ordered
states, and no local minima either. If $t'=0$ then 
delocalisation takes place necessarily through the 
magnetic sites and the deep minima in configuration space
represent ordered states while the `grassy' features
indicate shallow metastable states close to them. Our MC
data probing AF states at $t'=0$ suggests this picture,
curve (a) in Fig.18. At intermediate $t'$ the ordered states
are shallower, but the metastable states seem to have been
affected even more drastically, if our MC data, Fig.11, is
to be taken seriously. 

While the discussion above seems to be merely an analysis of
trends in the MC annealing process, a simpler energy 
landscape would make the occurence of AF states more likely
in the real materials as well.

\begin{figure}[t]
\psfig{figure=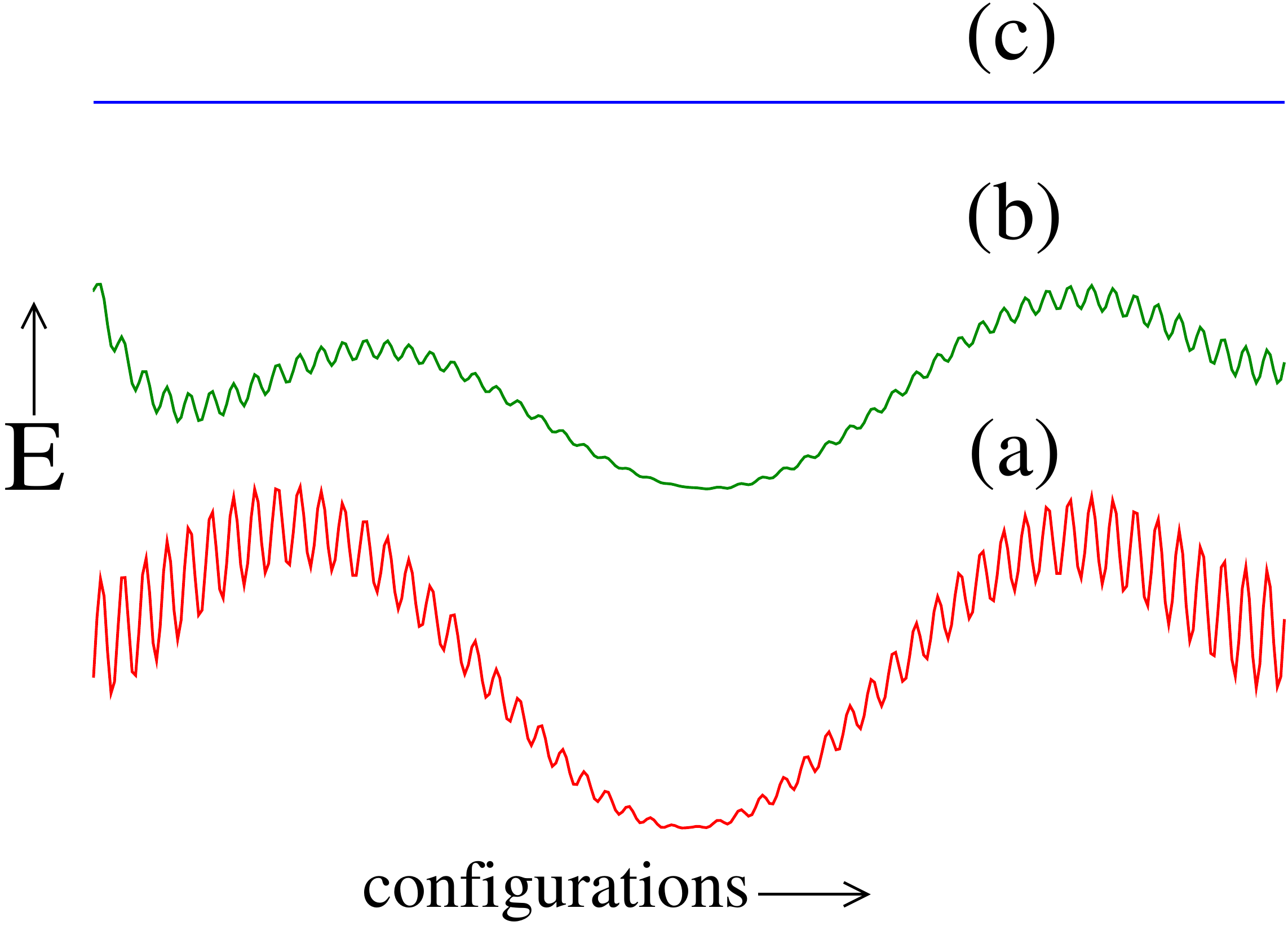,width=6.0cm,height=6.5cm,angle=0}
\caption
{\label{lndscp} Colour online: A schematic energy landscape of the
double perovskite model, in the space of spin configurations, for
(a).~$t'/t=0$, (b).~$t'/t \sim {\cal O}(1)$, and (c).~$t'/t
\rightarrow \infty$.
The panels are indicated for a fixed set of electronic parameters,
except $t'$ which varies as indicated above. At $t'=0$ the landscape
has many local minima around the deep minimum, and while the `binding
energy' of the ordered state, with respect to the paramagnet, is
large (and so also the $T_c$) the system is apt to get stuck in one
of the neighbouring minima in the cooling process. When $t'/t
\rightarrow \infty$ magnetic order makes no difference to the energy,
the electrons bypass the magnetic sites. At intermediate $t'$, while
the binding energy and $T_c$ are smaller, the local minima also seem
to be fewer and shallower. This makes the ordered state easier to access.
}
\end{figure}

\section{Discussion}
\label{discuss}

The real double perovskites are multiband materials, involving
additional interaction effects and antisite disorder beyond what
we have considered in this paper. We feel it is necessary to 
understand in detail the phase diagram of the `simple' model 
we have studied, and then move to more realistic situations.  
Below, we first provide 
a qualitative comparison of the trends we observe
with experimental data, and then move to a discussion of
additional interactions.

There are no clear experimental signatures of metallic AF phases
yet, driven by the kind of mechanism that we have discussed.
So, the comparison to experiments is, at the moment,
confined to the $T_c$ scales \cite{dp-tc1,dp-tc2}
{\it etc}, of the ferromagnetic
DP's. 
In a material like SFMO the electron density can be
increased by doping La for Sr, {\it i.e}, 
compositions like Sr$_{2-x}$La$_x$FeMoO$_6$.
This was tried \cite{dp-tc1} and the $T_c$ increased
from 420K at $x=0$ to $\sim$ 490K at  $x=1$.
SFMO has threefold degeneracy of the active,
$t_{2g}$, orbitals while we have considered a one band
model. When we create a correspondence by dividing the
electron count by the maximum possible per unit cell
(3 in our case, 9 in the real material), in our units
$x=0$ corresponds to $n=0.33$ and $x=1$ to $n=0.66$.

When $t'=0$, as a function of $n$ the  $T_c$ 
peaks around $n=0.2$, Fig.8, quite far from the experimental
value. However, in the presence of $t'=-0.3$
and $\Delta=4$, Fig.16, the peak occurs above $n=1$.
So, modest $t'$ can generate the ferromagnetic window
that is observed, and produce a $T_c \sim 0.1t$.
For $t=0.5$eV, this is in the right ballpark.

Unfortunately, attempts to increase $n$ via
A site substitution also brings in greater
antisite disorder (B-B$'$ interchange) and
even the possibility of newer patterns of A site
ordering (!) complicating the analysis.
For example, one would try 
compositions of the form:
A$_{2-x}$A$'_x$BB$'$O$_6$, where A and A$'$ have 
different valence in an attempt to change $n$.
The assumption is that the A$'$ only changes $n$
without affecting other electronic parameters,
{\it i.e}, A$'$ ions do not order and remain in 
an alloy pattern. This may not be true.
In fact, at $x=1$, the material 
AA$'$BB$'$O$_6$ may have a specific A-A$'$-B-B$'$ 
ordering pattern that affects electronic parameters
in a non trivial way and one cannot understand 
this material as a perturbation on A$_{2}$BB$'$O$_6$.
In such a situation one needs guidance from 
experiments and {\it ab initio} theory to fix
electronic parameters as $x$ is varied. All this
before one even considers the inevitable antisite
(B-B$'$) disorder and its impact on magnetism 
\cite{vn-pm-af-asd}.

\section{Conclusions}
\label{conclude}

We have studied a one band model of double perovskites in three
dimensions in the limit of strong electron-spin coupling on the
magnetic site. The magnetic lattice in the cubic double perovskites
is FCC and increasing the electron density leads from the
ferromagnet, through A and C type collinear antiferromagnets, 
to spiral or `flux' phases close to half-filling. We estimate the
$T_c$ of these phases, via Monte Carlo and variational calculation,
and find the AF $T_c$ to be significantly suppressed compared to the
2D case. We attribute it to the geometric frustration on the FCC 
lattice. The introduction of B$'$B$'$ hopping $t'/t \sim 0.3$
significantly alters the phase diagram and $T_c$ scales and 
creates a closer correspondence to the experimental situation on
DP ferromagnets.

\begin{acknowledgments}
We acknowledge use of the high performance computing
facility at HRI. PM was supported by a DAE-SRC Outstanding 
Research Investigator Award,
and the DST India via the Indo-EU ATHENA project.
\end{acknowledgments}
\vspace{.2cm}

%\section{Appendix}
\appendix
\section{}
\label{appdx}
Here we show how to calculate dispersion for selected ordered phases, which have
relatively small unit cells. We define the unit cell for each phase, and go to
$k-$space where the hamiltonian becomes block diagonal.

\subsection{Spectrum for collinear phases}

The Hamiltonian can be diagonalized by Fourier transformation
We write the
Hamiltonian $H$ as $H = H_{0}+H_{J}$,
where, $H_0$ is given by,
\begin{eqnarray}
  \nonumber
  H_{0} =\sum_{\vec{X},\sigma} [ \epsilon_{1}f^{\dagger}_{\vec{X},\sigma}f_{\vec{X},\sigma}
  +\epsilon_{2}m^{\dagger}_{\vec{X}+\vec{a_1},\sigma}m^{\dagger}_{\vec{X}+\vec{a_1},\sigma}]\\\nonumber
  - t'\sum_{\vec{X},\sigma}\sum_{\vec{\delta}\in \textrm{NNN}}(m^{\dagger}_{\vec{X}+\vec{a_1},\sigma}m_{\vec{X}
    +\vec{a_1}+\vec{\delta},\sigma}+\textrm{h.c.})\\
  - t\sum_{\vec{X},\sigma}\sum_{\vec{\delta} \in \textrm{NN}} (f^{\dagger}_{\vec{X},\sigma}m_{\vec{X}+\vec{\delta},\sigma}+\textrm{h.c.})
\end{eqnarray}
and $H_J$ is given by
\begin{equation}
  H_{J} = J \sum_{\vec{X}}\vec{S}(\vec{X}) \cdot \vec{\sigma}_{\alpha,\beta}f^{\dagger}_{\vec{X},\alpha}f_{\vec{X},\beta}
\end{equation}

\begin{figure}[b]
  \begin{center}
    \psfig{figure=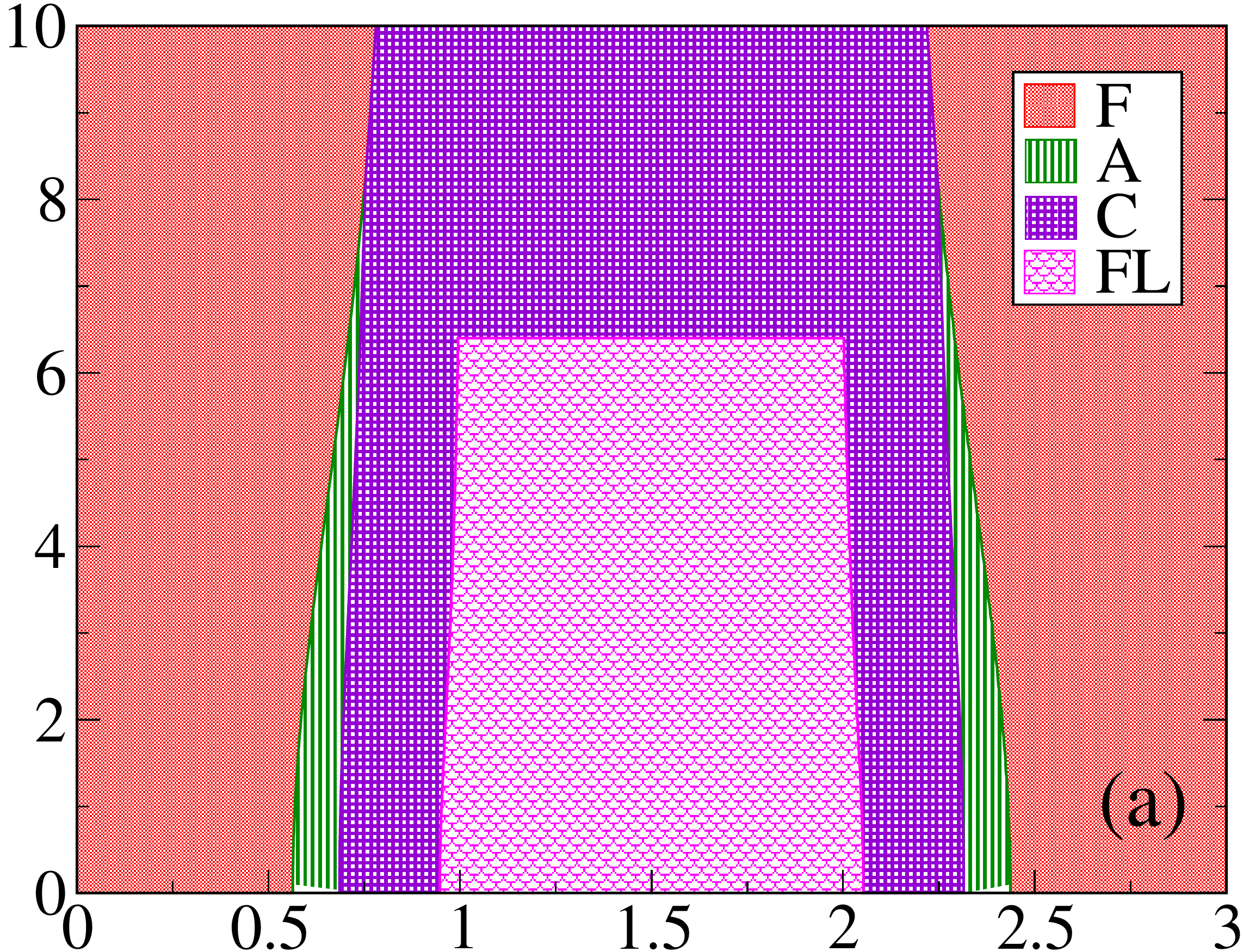,width=5.5cm,height=4cm,angle=0}
    \psfig{figure=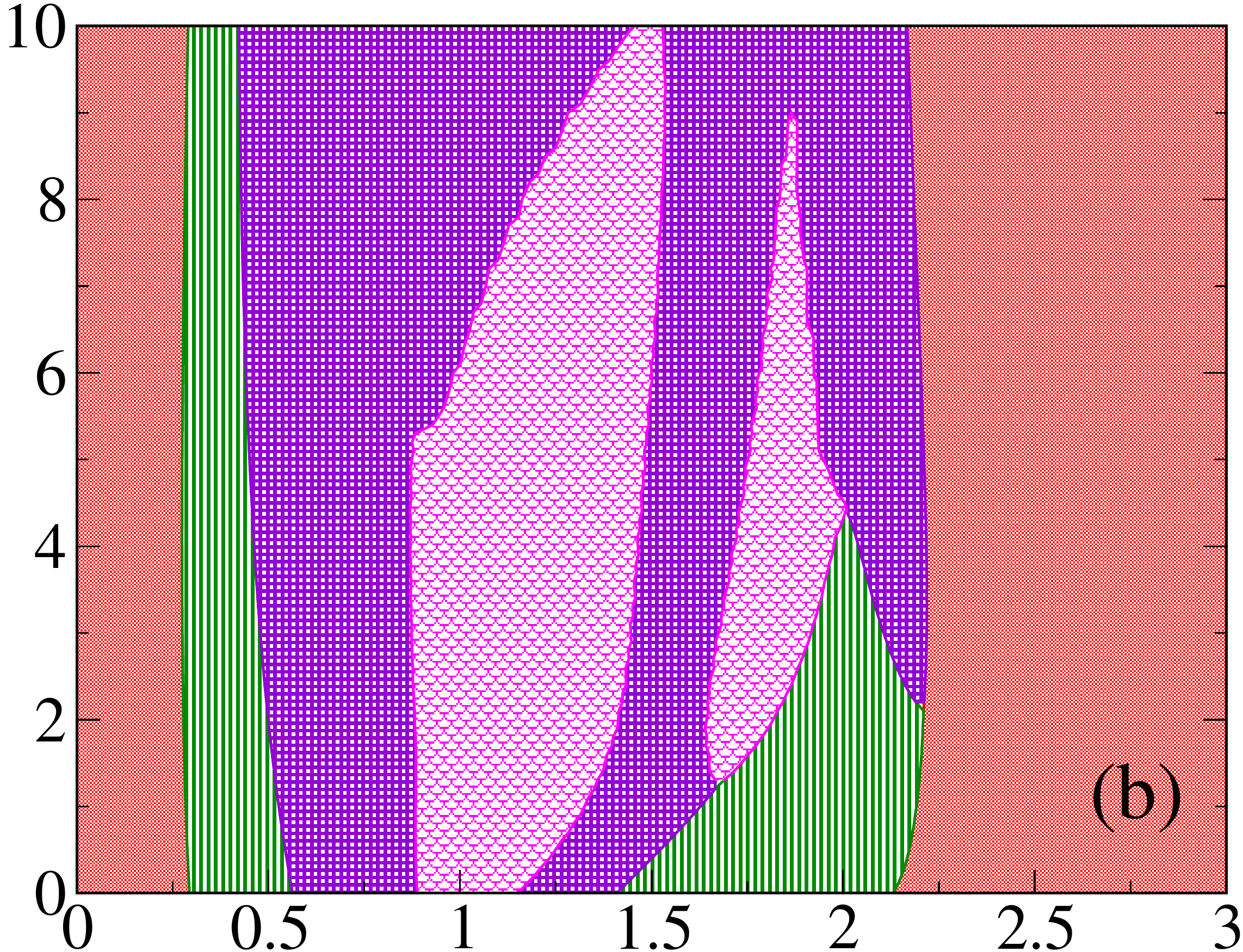,width=5.5cm,height=4cm,angle=0}
    \psfig{figure=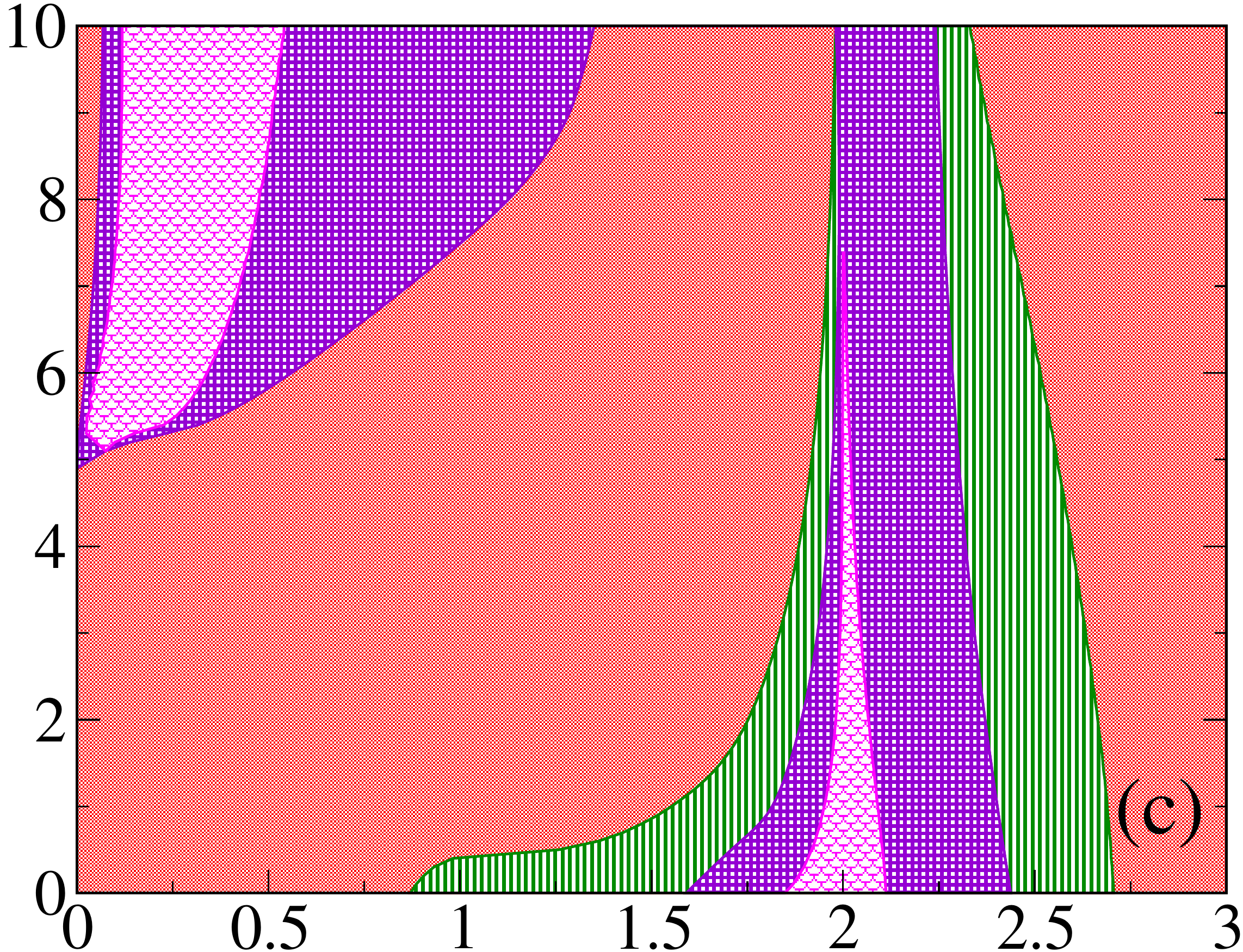,width=5.5cm,height=4cm,angle=0}
  \end{center}
  \caption{\label{phdk} Colour online: Phase diagram based on $k$ space
based diagonalisation for $t'=0,0.3,-0.3$. System size
 $N=160^{3}$. Here we can only use  F, A, C and `flux' phase as candidate
states but some of the complexity of more eleborate phase diagrams,
Fig.2, are already present.
  }
\end{figure}

The lattice vector $\vec{X}$ is defined as $\vec{X} = n_{1}\vec{A_{1}}
+ n_{2}\vec{A_{2}} + n_{3}\vec{A_{3}}$ with $A_i,i=1,2,3$ as the primitive
lattice vectors ($A_1=(2,0,0)$, $A_2=(1,1,0)$, $A_3=(0,1,1)$), defining
the periodicity of lattice with the 2 site unit cell. With this periodicity,
the unit cell has one "f" and one "m" site at $(0,0,0)$ and $(1,0,0)$
respectively. Now doing a Fourier transform on "$f$" operators (similarly for "$m$"s)
\begin{eqnarray}
  \label{eq:ft}
  f^{\dagger}_{\vec{X},\sigma} = \frac{1}{\sqrt{N}}\sum_{\vec{k}}f^{\dagger}_{\vec{k},\sigma}\exp(i\vec{k}\cdot\vec{X})\\
  f_{\vec{X},\sigma} = \frac{1}{\sqrt{N}}\sum_{\vec{k}}f_{\vec{k},\sigma}\exp(-i\vec{k}\cdot\vec{X})
\end{eqnarray}
This simplifies the non-magnetic part $H_{0}$ as follows,
\begin{eqnarray}\nonumber
  H_{0} = \sum_{\vec{k},\sigma} [ (\epsilon_{2}+A'_{\vec{k}})m^{\dagger}_{\vec{k},\sigma}m_{\vec{k},\sigma}\qquad\\
  +\epsilon_{1}f^{\dagger}_{\vec{k},\sigma}f_{\vec{k},\sigma}+(A_{\vec{k}}f^{\dagger}_{\vec{k},\sigma}m_{\vec{k},\sigma}+\textrm{h.c.})]
\end{eqnarray}
\begin{equation}
  =\sum_{\vec{k},\sigma}\left(f^{\dagger}_{\vec{k},\sigma} m^{\dagger}_{\vec{k},\sigma} \right)
  \left(\begin{array}{cc} \eps_1 & A_{\vec{k}}\\A_{\vec{k}} & \eps_2+A^{'}_{\vec{k}} \end{array}\right)
  \left(\begin{array}{c} f_{\vec{k},\sigma} \\ m_{\vec{k},\sigma} \end{array}\right)
\end{equation}
Which is reduced to $2\times 2$ block. the amplitudes $A_{\vec{k}}$ and $A'_{\vec{k}}$ are just
the cubic and FCC dispersions given by
\begin{eqnarray}\nonumber
  A_{\vec{k}}=-2t(\cos{k_{x}}+\cos{k_{y}}+\cos{k_{z}})\qquad\textrm{and}\\\nonumber
  A'_{\vec{k}} = -4t'(\cos{k_{x}}\cos{k_{y}}+\cos{k_{y}}\cos{k_{z}}+\cos{k_{z}}\cos{k_{x}})
\end{eqnarray}
Next, we have to simplify the $H_J$ part. For the collinear phases,
$\vec{S}(\vec{X})$ can be expressed as
\begin{equation}
  \vec{S}(\vec{X}) = 
  \left( \begin{array}{c} 0 \\ 0 \\ 1 \end{array}\right)
  \exp\left(i\vec{q}\cdot\vec{X}\right)
\end{equation}
For FM, $\vec{q}$ is trivially $(0,0,0)$. For A-type, $\vec{q} =
(\frac{\pi}{2},-\frac{\pi}{2},\frac{\pi}{2})$, while for C-type
$\vec{q} = (0,\pi,-\pi)$.
Now, plugging this value of $\vec{S}(\vec{X})$ in $H_J$ and doing
the Fourier transform for the $H_J$, we get,
\begin{equation}
  H_J = J\sum_{\vec{x}}\sigma
  f^{\dagger}_{\vec{k},\sigma}f_{\vec{k}+\vec{q},\sigma} \quad ;\sigma= \pm 1
\end{equation}

Now $\vec{q}=0$ for FM, so $H_J$ becomes diagonal. Thus total
hamiltonian $H$ still remains $2\times 2$ block, and the eigenvalues
for the FM are solutions of the following $2\times 2$ block
\begin{equation}
  \label{eq:fmblk}
  H_{2X2}(\vec{k},\sigma) = \left(
    \begin{array}{cc}
      \epsilon_1 + J\sigma& A_{\vec{k}} \\
      A_{\vec{k}} & \epsilon_{2} + A'_{\vec{k}}
    \end{array}
  \right)
\end{equation}

For A-type and C-type phases, we get matrix elements connecting
$|\vec{k},\sigma\rangle \to |\vec{k}+\vec{q},\sigma\rangle \to |\vec{k},\sigma\rangle$
, so that now we get to solve following $4\times4$ block
\begin{equation}
  \label{eq:acblk}
  H_{4\times4}(\vec{k},\sigma) =
  \left(
    \begin{array}{cccc}
      \epsilon_1 & J\sigma            & A_{\vec{k}} &          0         \\
      J\sigma    & \epsilon_1         &     0      & A_{\vec{k}+\vec{q}} \\
      A_{\vec{k}}  &     0              & \epsilon_2+ A'_{\vec{k}} & 0  \\
      0          & A_{\vec{k}+\vec{q}} &     0      & \epsilon_2+ A'_{\vec{k}}
    \end{array}
  \right)
\end{equation}

From these we obtain the spectrum for F,A,C phases on large ($\sim 100^3-500^3$) lattices,
which can be used to calculate the density of states, phase diagram, phase separation
windows etc.

\subsection{Spectrum for the `flux' phase}
The unit cell for the `flux' phase has 4$B$, and 4$B'$ atoms lying on the corners of the cube.
\begin{wrapfigure}{r}{0.1\textwidth}
  \vspace{-10pt}  \hspace{-10pt}
  \psfig{figure=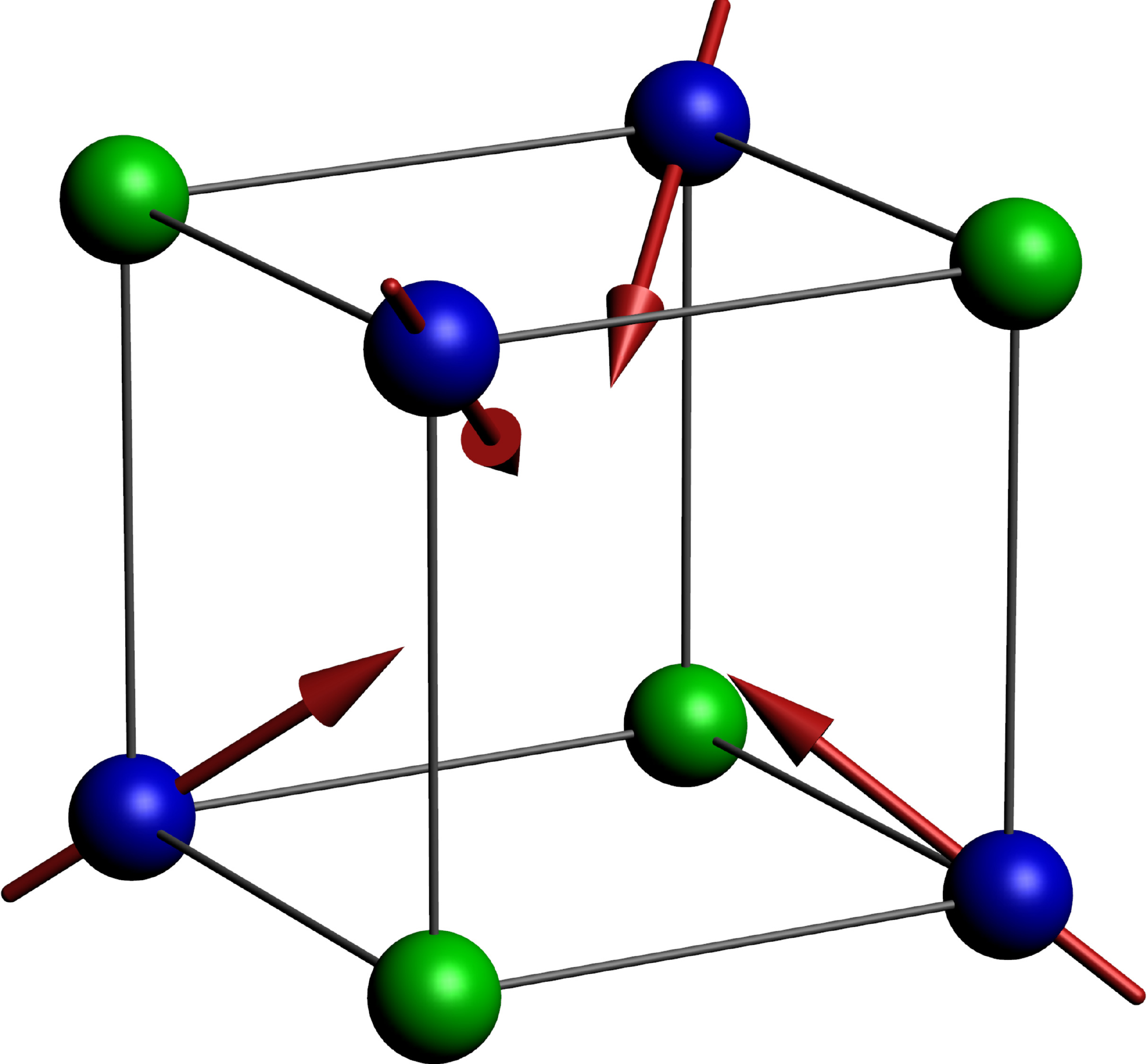,width=0.1\textwidth}  
  \vspace{-10pt}
\end{wrapfigure}
The primitive lattice vectors become $A_i=\{(2,0,0),(0,2,0),(0,0,2)\}$
At finite $J$, the same procedure (as for collinear phases) will reduce the hamiltonian
into $16\times 16$ block. To make life a bit simple, we use the J$\rightarrow \infty$ limit
on the hamiltonian for the `flux' phase, which is same as used in~\cite{sanyal:054411} except
its the 3D version.

This gives us 4 spinless $f_i$ levels and 8 $m_{i,\sigma}$ levels in the unit cell, which upon
simplification reduces to $12\times 12$ block.
With the basis $\left(\begin{array}{cc} f_i(k) & m_{i\uparrow}(k) \end{array}\right)_{i\in \{1,2,3,4\}}$
The hamiltonian breaks into $12\times 12$ block given as follows
% The reciprocal vectors $(\pi,0,0)$,$(0,\pi,0)$,$(0,0,\pi)$ define the brillouine zone,
% which is $\frac{1}{8}$th of the cubic system.
\begin{widetext}
  \begin{equation}\nonumber
    H = 
    \left(
      \begin{array}{cccccccccccc}
        \Delta & 0 & 0 & 0 & t_{1\uparrow}a_1 & t_{1\downarrow}a_1 & t_{1\uparrow}a_2 & t_{1\downarrow}a_2 & 0 & 0 & t_{1\uparrow}a_3 & t_{1\downarrow}a_3 \\
        0 & \Delta & 0 & 0 & t_{2\uparrow}a_2 & t_{2\downarrow}a_2 & t_{2\uparrow}a_1 & t_{2\downarrow}a_1 & t_{2\uparrow}a_3 & t_{2\downarrow}a_3 & 0 & 0 \\
        0 & 0 & \Delta & 0 & t_{3\uparrow}a_3 & t_{3\downarrow}a_3 & 0 & 0 & t_{3\uparrow}a_2 & t_{3\downarrow}a_2 & t_{3\uparrow}a_1 & t_{3\downarrow}a_1 \\
        0 & 0 & 0 & \Delta & 0 & 0 & t_{4\uparrow}a_3 & t_{4\downarrow}a_3 & t_{4\uparrow}a_1 & t_{4\downarrow}a_1 & t_{4\uparrow}a_2 & t_{4\downarrow}a_2 \\
        t_{1\uparrow}^{*}a_1 & t_{2\uparrow}^{*}a_2 & t_{3\uparrow}^{*}a_3 & 0 & 0 & 0 & t_{12} & 0 & t_{23} & 0 & t_{13} & 0 \\
        t_{1\downarrow}^{*}a_1 & t_{2\downarrow}^{*}a_2 & t_{3\downarrow}^{*}a_3 & 0 & 0 & 0 & 0 & t_{12} & 0 & t_{23} & 0 & t_{13} \\
        t_{1\uparrow}^{*}a_2 & t_{2\uparrow}^{*}a_1 & 0 & t_{4\uparrow}^{*}a_3 & t_{12} & 0 & 0 & 0 & t_{13} & 0 & t_{23} & 0 \\
        t_{1\downarrow}^{*}a_2 & t_{2\downarrow}^{*}a_1 & 0 & t_{4\downarrow}^{*}a_3 & 0 & t_{12} & 0 & 0 & 0 & t_{13} & 0 & t_{23} \\
        0 & t_{2\uparrow}^{*}a_3 & t_{3\uparrow}^{*}a_2 & t_{4\uparrow}^{*}a_1 & t_{23} & 0 & t_{13} & 0 & 0 & 0 & t_{12} & 0 \\
        0 & t_{2\downarrow}^{*}a_3 & t_{3\downarrow}^{*}a_2 & t_{4\downarrow}^{*}a_1 & 0 & t_{23} & 0 & t_{13} & 0 & 0 & 0 & t_{12} \\
        t_{1\uparrow}^{*}a_3 & 0 & t_{3\uparrow}^{*}a_1 & t_{4\uparrow}^{*}a_2 & t_{13} & 0 & t_{23} & 0 & t_{12} & 0 & 0 & 0 \\
        t_{1\downarrow}^{*}a_3 & 0 & t_{3\downarrow}^{*}a_1 & t_{4\downarrow}^{*}a_2 & 0 & t_{13} & 0 & t_{23} & 0 & t_{12} & 0 & 0
      \end{array}
    \right)
  \end{equation}
  \begin{eqnarray}\nonumber
    \textrm{Where the symbols in the above are defined as }\quad
    \alpha = \sqrt{\frac{\sqrt{3}+1}{2\sqrt{3}}};\quad \beta = \sqrt{\frac{\sqrt{3}-1}{2\sqrt{3}}};\quad z = \frac{1-i}{\sqrt{2}}\\\nonumber
    a_1 = 2\cos{k_1};\quad  a_2 = 2\cos{k_2};\quad  a_3 = 2\cos{k_3};\quad
    t_{12} = -4t'\cos{k_1}\cos{k_2};\quad t_{23} = -4t'\cos{k_2}\cos{k_3};\\\nonumber
    t_{13} = -4t'\cos{k_1}\cos{k_3};\quad  t_{1\uparrow} = t_{2\uparrow} = -t\alpha;\quad t_{3\uparrow} = t_{4\uparrow} = -t\beta;\quad 
    t_{1\downarrow} = -t_{2\downarrow} = tz\beta;\quad t_{3\downarrow} = -t_{4\downarrow} = -tz^{*}\alpha
  \end{eqnarray}
\end{widetext}

\bibliographystyle{unsrt}

\begin{thebibliography}{99}
\bibitem{footnote1}
  We have worked on $\Delta \in [-10,10]$,
  so J has to be large compaired to it. 
  We used J = 1000, as the atomic bands which,
  for finite J, are actually dispersive (of band-width 
  $\sim {\cal{O}}( \frac{1} {\Delta} )$) become
  actually atomic to required numerical accuracy.
\bibitem{footnote2}
  We campared ratio of the maximum of $(\delta E)^{FM}$
  and maximum of $(\delta E)^{AF}$ in the density window,
  for $\Delta = 0$ and $t' = 0$
\end{thebibliography}

\end{document}